\documentclass[aps,prd,twocolumn,preprintnumbers,nofootinbib]{revtex4-1}

\usepackage{graphicx}
\usepackage[FIGTOPCAP]{subfigure}
\usepackage{dcolumn}
\usepackage{bm}
\usepackage{amsmath}
\usepackage{epstopdf}
\usepackage{amsmath,amssymb}
\usepackage{bigints}
\usepackage{mathtools}
\usepackage{array}
\usepackage{bbm}
\usepackage{dsfont}
\usepackage{tikz}
\usepackage{tikz-3dplot}
\usepackage[normalem]{ulem} 
\usepackage[breaklinks=true,colorlinks,citecolor=blue,linkcolor=blue,urlcolor=blue]{hyperref}
\usepackage{appendix}

\newcommand{\beq}{\begin{eqnarray}}
\newcommand{\eeq}{\end{eqnarray}}

\newcommand{\real}{{\sf I}\kern-.12em{\sf R}}
\newcommand{\comp}{{\sf I}\kern-.50em{\sf C}}
\newcommand{\unity}{{\sf I}\kern-.54em{\sf 1}}

\newcommand{\norm}[1]{\left\lVert#1\right\rVert}

\usepackage{xargs}                      

\begin{document}

\title{Spectral Analysis of Causal Dynamical Triangulations via Finite Element Method}

\author{Fabio Caceffo}
\email{fabio.caceffo@phd.unipi.it}
\affiliation{Dipartimento di Fisica dell'Universit\`a di Pisa and INFN
	- Sezione di Pisa,\\ Largo Pontecorvo 3, I-56127 Pisa, Italy.}

\author{Giuseppe Clemente}
\email{giuseppe.clemente@desy.de}
\affiliation{Dipartimento di Fisica dell'Universit\`a di Pisa and INFN
	- Sezione di Pisa,\\ Largo Pontecorvo 3, I-56127 Pisa, Italy.}
\affiliation{Deutsches Elektronen-Synchrotron (DESY), Platanenallee 6, 15738 Zeuthen, Germany}


\begin{abstract}

    We examine the dual graph representation of simplicial manifolds in
    Causal Dynamical Triangulations (CDT) as a mean to build 
    observables, 
    and propose a new representation based on the Finite Element Methods (FEM).
    In particular, 
    with the application of FEM techniques, 
    we extract the (low-lying) spectrum of the Laplace-Beltrami (LB) operator on the 
    Sobolev space $H^1$ of scalar functions on piecewise flat manifolds, 
    and compare them with corresponding results obtained by using the dual graph representation.
    We show that, besides 
    for non-pathological cases in two dimensions, 
    the dual graph spectrum and spectral dimension 
    do not generally agree, neither quantitatively nor qualitatively, 
    with the ones obtained from the LB operator on the continuous space. 
    We analyze the reasons of this discrepancy and discuss its possible implications 
    on the definition of generic observables built from the dual graph representation. 

\end{abstract} 

\maketitle

\section{Introduction}

The search for a reliable set of observables to 
probe the geometry of simplicial manifolds, allowing 
the exploration and characterization of the phase diagram, 
has become one of the central pursues 
of the Causal Dynamical Triangulations (CDT)~\cite{cdt_report,cdt_review19} program of Quantum Gravity in recent years.

This has led to the introduction of new observables and techniques, 
many of which are based on the dual graph representation 
of simplicial manifolds, where geometric information is encoded in the 
adjacency relations between the elementary units of volume, called simplices.
To mention a few, Hausdorff~\cite{precdt_ambjorn_hausdorff,cdt_ambjorn_hausdorff_spectral} 
and spectral dimensions~\cite{edt_spectral_dim,diffproc,cdt_spectral_dim} 
are built from processes taking place on the graph dual to CDT triangulations, 
but also other quantities like the spectrum of the Laplace matrix of graphs dual 
to CDT spatial slices~\cite{lbstruct,lbrunning}, and the recently proposed 
quantum Ricci curvature~\cite{cdt_quantum_Ricci_curv,cdt_quantum_Ricci_curv_round}, 
are based on dual graphs constructions.

These observables have been proven to be unquestionably valuable 
in capturing some relevant geometric properties of simplicial manifolds.
In particular, observables built from the dual graph representation are given 
    the same geometric intepretation as the ones defined using the space of 
    functions on the piecewise continuous manifolds. 
    Therefore, as the main goal of the present work, we find interesting to investigate how these graph-theoretical representations
compare with the representations of observables on the same 
simplicial manifolds from which they are built from, at least at the larger scales.

In order to proceed in this direction,
we propose the family of Finite Element Methods (FEM), which, 
besides being backed by well-grounded 
mathematical framework~\cite{fem_allairebook,fem_hughesbook,fem_strangbook,fem_taylorbook,fem_babuska}, 
in its generality, allows also to properly represent things like local observables, 
coupling terms with other fields or with higher derivative metric terms in the action.
Finite element methods are not new in physics, 
and have been employed to model a huge variety of circumstances: 
indeed, FEM is one of the main tools of multiphysics 
simulations~\cite{fem_multiphysics_modeling,fem_analapps}, 
and found also recent applications in 
lattice quantum field theory~\cite{fem_Brower_phi4,fem_Brower_phi4_II}.
The FEM framework has many similarities with the Discrete Exterior Calculus (DEC),
which has been recently investigated in CDT to study approximate 
Killing symmetries~\cite{Brunekreef:2020red} and tensorial Laplacian spectra~\cite{Reitz:2022dbj}.

Despite the many possibilities which can be explored
(we just mention one of them in Appendix~\ref{sec:curvobs}),
in this work we mainly set the stage for future FEM studies 
by treating a specific problem:
the Laplace-Beltrami eigenvalue problem on CDT simplicial manifolds. 
In particular, after a presentation of the basics of the FEM formalism, 
we examine its behavior in some test cases,
and also compare new FEM results with
the earlier results of spectral analysis on dual graphs presented in~\cite{lbstruct,lbrunning} 
and reviewed in~\cite{lbpos19}. What we find, besides the expected convergence behavior of the FEM results to the spectrum of the Laplace-Beltrami (LB) operator on the simplicial manifold, is a disagreement with the dual graph method from a quantitative point of view: we try to explain the reason why this happens in Section~\ref{subsec:supr} by studying a highly pathological situation, in which the dual graph method fails to detect even some qualitative large-scale features of the manifold. This motivates us to take again in careful consideration the results obtained 
via dual graph methods found in literature, starting from the spectral ones.

The structure of the paper is the following.\\
In Section~\ref{sec:LBspectrum_geom} we review some general, representation independent, 
concept about the spectrum of the Laplace-Beltrami operator and its relation 
with geometrical properties of manifolds; in Section~\ref{subsec:lap_dg} we also 
explicitly show how the Laplace matrix of a dual graph approximates 
the Laplace-Beltrami operator of a simplicial manifold, since it proves helpful in discussing some specific features of the dual graph representation.\\
The Finite Element Method formalism is briefly introduced in Section~\ref{sec:femintro},
leaving a detailed description of its application to the solution of the Laplace-Beltrami
eigenvalue problem on simplicial manifolds to Appendix~\ref{sec:details}. \\
In Section~\ref{sec:examples} we consider some test geometries, for which we compare the spectra of dual graph and FEM representations, 
showing the convergence of the latter to the spectrum of the LB differential operator, 
and addressing the reasons why the first, instead, exhibits noticeable discrepancies.\\
Numerical results are shown in Section~\ref{sec:numres} where we compare the FEM results 
with the earlier ones obtained by using the dual graph representation.\\
Finally, in Section~\ref{sec:conclusions} we conclude by give some remarks about 
the dual graph and the FEM formalism, and discussing future perspectives.

\section{Useful relations between spectrum and geometry}\label{sec:LBspectrum_geom}

The Laplace-Beltrami (LB) operator $(-)D_\mu D^\mu$ can provide us very useful
information on the geometry of a manifold\footnote{In this work, we implicitly deal with 
    the Laplace-Beltrami operators acting on scalar functions only. 
    Generalization to tensor fields are possible, 
    and should be considered if one wants to introduce 
    consistently matter and gauge couplings.}.
Currently, two ways of expressing the relationship between the geometric properties 
of a Riemannian manifold and the spectrum of the Laplace-Beltrami operator associated 
to its metric have proved useful:
through the properties of diffusion processes 
on the manifold and by studying the (cumulative) 
density of ``energy levels'' of the manifold, as we review in Sections~\ref{subsec:difft} and~\ref{subsec:enelevden}. 
Both lead to the identification 
of typical (length/energy) scales of the manifold and a suitable definition 
of a scale-dependent ``spectral dimension''.

In order to study the spectral properties of simplicial manifolds 
corresponding to CDT configurations, some form of approximation of the spectrum 
of the LB operator is necessary In this respect, 
a careful assessment of the accuracy of the approximation used is then of utmost importance,
since the relations with the geometrical
properties of the manifold involve the \textit{spectrum of the LB differential operator},
which is defined on an infinite-dimensional space of functions.
In previous works, the spectrum of the LB operator has been approximated
with the spectrum of the \textit{Laplace matrix} 
of the \textit{dual graph} associated with the triangulation 
(i.e., the graph of the connections between the centers of adjacent simplices),
as discussed in Section~\ref{subsec:lap_dg}.
\subsection{Diffusion processes}\label{subsec:difft}

Following~\cite{cdt_report}, let us consider a diffusion process 
on a boundaryless manifold $\mathcal{M}$, described by the heat equation:
\begin{equation}\label{eq:heat}
\partial_t u(\mathbf{x},t)= D^\mu D_\mu u(\mathbf{x},t);
\end{equation}
and write $u(\mathbf{x},t)$ on the basis of the eigenfunctions 
of the Laplace matrix for each $t$:
\begin{equation}
u(\mathbf{x},t)=\sum_{n=0}^{\infty} c_n(t)f_n(\mathbf{x});
\end{equation}
then substituting into Equation~\eqref{eq:heat}, one obtains for the coefficients $c_n(t)$:
\begin{equation}
\frac{d}{dt} c_n(t)=-\lambda_n c_n(t) \quad \longrightarrow \quad c_n(t)=c_n(0) \; e^{-\lambda_n t},
\end{equation}
with solution $u$ for the heat Equation~\eqref{eq:heat}:
\begin{equation}\label{eq:heatsol}
u(\mathbf{x},t)=\sum_{n=0}^{\infty}c_n(0)\;f_n(\mathbf{x})\;e^{-\lambda_n t},
\end{equation}
with the coefficients $c_n(0)$ fixed by the initial conditions of the diffusion process. 
A diffusion process on a manifold without boundary can be seen as a continuous-time 
stochastic process, as the integral of the density $u$ is conserved in time:
\begin{equation}
\begin{gathered}
\partial_t \int_{\mathcal{M}} d^dx \sqrt{g(\mathbf{x})} \; u(\mathbf{x},t) = \int_{\mathcal{M}} d^dx \sqrt{g(\mathbf{x})}\; D^\mu D_\mu u(\mathbf{x},t) =\\
= \int_{\mathcal{M}} d^dx \; \partial_\mu ( \sqrt{g(\mathbf{x})}\; g^{\mu \nu} \partial_\nu u(\mathbf{x},t)) = 0,
\end{gathered}
\end{equation}
where we have exploited a well-known expression for the LB operator applied to a scalar function in terms of the underlying metric.\\
Now let us consider a process starting from a density all concentrated at one point:
\begin{equation}
u(\mathbf{x},0)=\frac{1}{\sqrt{\det(g(\mathbf{x}))}} \delta^d(\mathbf{x} - \mathbf{x}_0).
\end{equation}
The solution to the diffusion process $u$ can be expanded~\cite{heatk} for small $t$ in the form 
(adding the argument $\mathbf{x}_0$ as the starting point for the process):
\begin{equation}\label{eq:hsolseries}
    u(\mathbf{x},\mathbf{x}_0,t) \sim \Big[ \frac{e^{-d_g^2(\mathbf{x},\mathbf{x}_0)/4t}}{t^{d/2}} \Big] \sum_{n=0}^{\infty} a_n(\mathbf{x},\mathbf{x}_0) \; t^n,
\end{equation}
where $d_g(\mathbf{x},\mathbf{x}_0)$ is the \emph{geodesic distance} between points $\mathbf{x}$
and $\mathbf{x}_0$. 
The term inside the square brackets reproduces the behavior of the diffusion process 
on a $d$-dimensional flat space. This can be expected from the observation that a smooth manifold 
is always almost flat on small enough scales, and small diffusion times 
are related to small length scales, since from the last equation we can see that 
the typical length scale of the diffusion process is $l\sim\sqrt{t}$.

A key related quantity is the average \textit{return probability}~\cite{fractals_havlinbook,edt_spectral_dim,cdt_spectral_dim,diffproc}, defined as:
\begin{equation}\label{eq:retpr}
P(t):=\frac{1}{V} \int d^d \mathbf{x} \sqrt{\det(g(\mathbf{x}))} \; u(\mathbf{x},\mathbf{x},t),
\end{equation}
with analogous expansion for small $t$:
\begin{equation}\label{eq:retprseries}
P(t) \approx \frac{1}{t^{d/2}} \sum_{n=0}^{\infty} A_n \; t^n,
\end{equation}
where the $A_n$ have been obtained by integrating the coefficients
$a_n(\mathbf{x},\mathbf{x})$ of Equation~\eqref{eq:hsolseries}, and are related to geometric quantities such as volume, curvature and diffeomorphism invariant scalars built from the Riemann tensor~\cite{heatk}.\\
For an infinite flat space, the return probability reads:
\begin{equation}
P(t)=\frac{1}{(4\pi t)^{d/2}},
\end{equation}
from which one can extract the dimension $d$ by computing the logarithmic derivative:
\begin{equation}\label{eq:diffspdim}
d=-2 \frac{d \log P(t)}{d \log t},
\end{equation}
which is constant in diffusion time.

Taking inspiration from these observations, 
it is possible to define with 
Equation~\eqref{eq:diffspdim} a diffusion-time-dependent (and therefore scale-dependent) 
\textit{spectral dimension} for 
generic manifolds~\cite{edt_spectral_dim,cdt_spectral_dim,diffproc}. 
From Equation~\eqref{eq:heatsol}, it can be seen that, for $t\gg \frac{1}{\lambda_1}$, 
the solution of the heat Equation~\eqref{eq:heat} approaches the constant mode 
(that relative to the null eigenvalue), then the spectral dimension approaches zero: 
this happens if the spectrum is discrete, as it is the case for compact manifolds 
(and unlike $\mathbb{R}^d$, for example), which appear point-like 
(i.e., zero dimensional) at scales much larger then their typical size. 
From the expansion in Equation~\eqref{eq:retprseries}, 
instead, it is apparent that for small scales (i.e., short diffusion times) 
the spectral dimension coincides with the geometric dimension of the manifold. 
So, interesting geometric information is obtained 
when this quantity is evaluated at intermediate diffusion times: 
a diffusion mode gives its main contribution to the spectral dimension 
as long as $t\lesssim \frac{1}{\lambda}$ 
(thus having a typical length scale $\frac{1}{\sqrt{\lambda}}$), 
and is in general exponentially suppressed as the diffusion time increases.

\subsection{Energy level density}\label{subsec:enelevden}

Another useful definition of dimension 
is based on the integrated spectral density $n(\lambda)$, 
defined as the number of eigenvalues below a certain threshold $s$:
\begin{equation}\label{eq:numcount}
    n(\lambda) = \sum_{\lambda_i\in \mathcal{S}} \theta(\lambda-\lambda_i),
\end{equation}
where $\mathcal{S}$ is the spectrum under consideration, 
and $\theta$ is the Heaviside step function.

An interesting result, involving $n(\lambda)$, is the so-called Weyl's theorem~\cite{weyl}: 
the LB operator on smooth compact manifolds is such that asymptotically 
(i.e., for large eigenvalues) the integrated spectral density behaves as:
\begin{equation}\label{eq:weylth}
n(\lambda) \sim \frac{\omega_d}{(2\pi)^d}V \lambda^{\frac{d}{2}},
\end{equation}
where $d$ is the chart dimension of the manifold 
and $\omega_d$ is the volume of the $d$-dimensional ball of unit radius. 
This relation can be interpreted as the dependence between the integrated density 
of energy levels of the manifold and the energy, 
as in Quantum Mechanics the LB operator represents the kinetic energy.

The asymptotic relation in Equation~\eqref{eq:weylth} 
actually holds only for the higher part of the spectrum; 
nonetheless, 
extending its range of validity turns out to be useful in 
describing manifolds whose dimensionality at some scale 
(corresponding to a certain region of the LB spectrum)
appears different from its chart (UV) dimension $d$,
leading us to the definition of 
another kind of effective spectral dimension, introduced in~\cite{lbstruct}:
\begin{equation}\label{eq:elevspdim}
\frac{2}{d_{EFF}} \equiv \frac{d \log \lambda}{d \log (n/V)}.
\end{equation}

In general, then, we are assuming that by studying the behavior of $\lambda$ 
as a function of $n/V$ we can detect the appearance of particular structures 
related to a specific dimensional behavior when 
the related length scale is reached. 
This length scale, let us refer to it as $l$, 
is related to $n/V$ by the relation $n/V\sim l^{-d}$, 
where $d$ is the chart dimension of the manifold, not only for dimensional reasons, 
but also because of this simple consideration: 
suppose that the low part of the spectrum 
has effective dimensionality $D<d$ and that some the other ``transverse'' 
dimensions $d-D$ have typical scale $l$, 
then the behavior of $n(\lambda)$ until the minimum value to excite the 
transverse dimensions of about $\pi^2/l^2$ is reached is something 
like $n(\lambda)=\omega_d(V/l^{d-D})\lambda^{D/2}$; 
substituting the turning-point value of $\lambda$ we obtain what is stated above. 
Some additional care with this relation is required when there are 
transverse dimensions with different associated typical scales.

\subsection{The Laplace matrix of the dual graph}\label{subsec:lap_dg}

\begin{figure}[h]
	\centering
	\includegraphics[width=\linewidth]{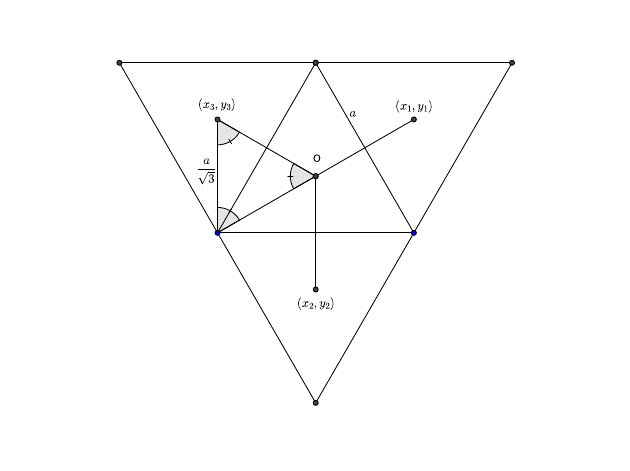}
	\caption{The construction to show the idea behind the approximation of the LB operator with the dual graph method.}
    \label{fig:dgcalc}
\end{figure}

The eigenproblem of the LB operator on the simplicial manifolds involved in CDT 
is not generally solvable by analytical means, 
thus a proper approximation technique is needed.
The method used in earlier spectral studies consists in substituting the real LB operator with
the Laplace matrix of the \textit{dual graph} associated with the triangulation,
that is, the graph obtained connecting the centers of the $d$-simplices 
sharing a $(d-1)$-simplex, and is based on the following idea, 
exemplified for convenience in two dimensions:
let us consider an \textit{equilateral} triangle with sides $a$ 
and center in the origin $O$ of a cartesian plane, 
adjacent to three other equilateral triangles, whose centers have 
coordinates $\{(x_i,y_i)\}_{i=1,2,3}$ respectively;
these points are at distance $a/\sqrt{3}$ from the origin, 
as depicted in Figure~\ref{fig:dgcalc}.

The Taylor approximation to second order in $a$ around $O$ for a function $f$
evaluated at the points $\{(x_i,y_i)\}_{i=1,2,3}$ reads:
{\small
\begin{equation}\label{eq:taylor}
\begin{gathered}
\forall i=1,2,3\;\;\;\; f(x_i,y_i)=f(0,0)+(x_i,y_i) \cdot \nabla f (0,0) +\\
+\frac{1}{2}\; [\partial_x^2 f(0,0) x_i^2+\partial_y^2 f(0,0) y_i^2+2\partial_x\partial_y f(0,0) x_i y_i] + o(a^3).\\
\end{gathered}
\end{equation}}
Then substituting the coordinates of the points:
{\small
\begin{equation}
\begin{gathered}
(x_1,y_1)=\frac{a}{\sqrt{3}}(\frac{\sqrt{3}}{2}, \frac{1}{2}),\\
(x_2,y_2)=\frac{a}{\sqrt{3}}(0,-1),\\
(x_3,y_3)=\frac{a}{\sqrt{3}}(-\frac{\sqrt{3}}{2}, \frac{1}{2}),
\end{gathered}
\end{equation}}
and summing the three Equations~\eqref{eq:taylor}, one obtains:
\begin{equation}
\scriptstyle
f(x_1,y_1)+f(x_2,y_2)+f(x_3,y_3) - 3f(0,0)=\frac{1}{4}a^2 \triangle f(0,0) + o(a^3),
\end{equation}
implying
\begin{equation}\label{eq:dgfinal}
\scriptstyle
\triangle f(0,0) = \frac{1}{a^2} \; 4 \; [f(x_1,y_1)+f(x_2,y_2)+f(x_3,y_3) - 3f(0,0)] + o(a).
\end{equation}

A similar relation holds for the center of each triangle of the simplicial manifold, 
so the LB operator is approximated through a matrix acting on a vector space 
of dimension equal to the number of triangles 
(the space of the possible values of the function in the centers): 
\begin{equation}
-\triangle \leftrightarrow L \; = \; 3 \cdot \mathds{1} - A,
\end{equation}
where $A$ is the adjacency matrix, whose entry $A_{ij}$ is $1$ if the triangles labeled 
with $i$ and $j$ are adjacent and $0$ otherwise. 
The factor $\frac{1}{a^2}$ is ignored since it is used as measure unit, 
while the factor $4$ is an overall scale factor, whose value will come in useful to
compare the results with the new method that we will introduce in the next Section.

In higher dimensions, analogous calculations, that are possible only 
for \textit{regular} (i.e., equilateral) simplices, lead us to approximate the Laplace-Betrami operator 
(forgetting for a while the overall constant) with the matrix:
\begin{equation}\label{eq:disclap}
-\triangle \rightarrow L\;= \; (d+1) \cdot \mathds{1} - A.
\end{equation}
It is straightforward, though quite tedious, to prove that in generic dimension $d$ 
the overall numeric constant appearing in the analogous of 
Equation~\eqref{eq:dgfinal} equals $d^2$.

The matrix $L$ in Equation~\eqref{eq:disclap} shares many properties with the LB operator:
it is symmetric and positive-semidefinite\footnote{The positive-semidefiniteness 
    of $L$ comes from its the fact that it has positive elements on the diagonal and it is 
    \textit{diagonally dominant}, i.e., 
    the absolute value of each diagonal element is greater or equal to the sum of the 
absolute values of the other terms in the same row}, with a unique eigenvector associated to 
the zero eigenvalue, the uniform function\footnote{The zero-mode is unique for manifolds with a 
    single connected component. In the case of the matrix $L$, it comes from the fact 
    that each row adds up to $0$, since each $d$-simplex has exactly $d+1$ neighbours}.
Moreover, $L$ is \textit{sparse} (most of its entries are $0$),
thus making it possible to numerically calculate its spectrum
using specifically optimized algorithms.

\section{Introduction to Finite Element Methods for spectral analysis}
\label{sec:femintro}

As apparent from the discussion in the previous section,
one reason that drove us to introduce an alternative method to
approximate the LB spectrum is that
a faithful correspondence between the spectrum of the Laplace matrix of the graph
dual to a triangulation and its exact Laplace-Beltrami spectrum 
requires, at least, the simplices to be regular, which is not the case 
for full CDT triangulations, where generic values of the parameter $\Delta$ encode the asymmetry
between spacelike and timelike links in the Euclidean space.

Moreover, for the method that we are going to introduce,
it is always possible to set up an iterative
procedure~(Appendix~\ref{subsec:refinement})
whose convergence to the exact spectrum of the 
LB operator is guaranteed by standard theoretical results in literature.\\
In the following, we outline the basic ideas the Finite Element Method relies upon, leaving the technical details of our application to CDT to Appendix~\ref{sec:details}.\\

Finite Element Methods (FEM) are a family of approximation techniques for the solution 
of partial derivative equations that is widely studied and applied in many fields 
where complex modeling is necessary~\cite{fem_allairebook,fem_hughesbook,fem_strangbook,fem_taylorbook,fem_babuska}. 
Complex objects are decomposed into simpler smaller parts to reduce 
the number of degrees of freedom to a finite one, which is far easier to be dealt with.\\
The application of FEM in the context of spectral analysis of manifolds
relies upon a weak formulation of the Laplace-Beltrami eigenproblem,  
which, on a (simplicial) manifold $\mathcal{M}$ without boundary
takes the form\footnote{The minus sign is a convention mainly adopted in mathematics, 
and that we follow since it makes the spectrum non-negative.}:
\begin{equation}\label{eq:eigp}
    -\triangle f(\mathbf{x})= \lambda f(\mathbf{x}).
\end{equation}
By multiplying both sides of this equation with an arbitrary test function $\phi$,
and integrating over the whole manifold, we obtain:
\begin{equation}\label{eq:weak-eigp}
\int_\mathcal{M}\! d^dx \;\nabla \phi(\mathbf{x}) \nabla f(\mathbf{x}) = \lambda \int_\mathcal{M}\! d^dx \; \phi(\mathbf{x}) f(\mathbf{x}),
\end{equation}
where a step of integration by parts has also been performed.

This second form, where $f$ is assumed to be 
a reasonable object (like a Sobolev function or a distribution), 
is equivalent to the one in Equation~\eqref{eq:eigp}. 
Nonetheless, it is useful to see the problem 
in this form because its natural environment of definition is wider.
The usual environment for these kinds of problems is the Sobolev space $H^1(\mathcal{M})$, 
the space of $L^2$ scalar functions that admit weak first derivatives, 
as it is a set of quite regular functions in which it can be proved that 
solutions exist for the problem.
In the following, we refer to the spectrum of the LB operator on the class 
of $H^1(\mathcal{M})$ scalar functions as the \emph{exact LB spectrum}.

FEM consists in solving a problem similar to the one shown 
in Equation~\eqref{eq:weak-eigp} in a sequence $\{\mathcal{V}_r\}_{r=0}^\infty$ 
of particular finite-dimensional 
subspaces of $H^1$ with increasing dimension $\mathcal{V}_r\to H^1$, 
whose eigenvectors $f_n^{(r)}$ and eigenvalues $\lambda_n^{(r)}$ converge to the exact LB 
eigenvectors and eigenvalues of the infinite-dimensional problem~\eqref{eq:weak-eigp} in $H^1$.
 
In these finite-dimensional subspaces, as outlined in Section~\ref{subsec:fineigp}, the problem consists in nothing but an eigenproblem of a finite-dimensional matrix, whose entries are calculated in Section~\ref{subsec:mat_elem}.\\
Because of the (quite natural) choices we make to define our sequence of subspaces for our simplicial manifolds, this progression can be seen as a series of subsequent \textit{refinements} of the starting triangulation: each step consists in subdividing every simplex of the triangulation into smaller ones, while preserving a simplicial manifold structure. 
For details, see Section~\ref{subsec:refinement}.

\section{Comparison between FEM and dual graph methods on test geometries}\label{sec:examples}

Before delving into the application of FEM to real CDT cases, 
it is useful to investigate its behaviour in some simple exemplar situations.
This, besides providing us with some checks on the expected convergence of the method 
to the exact LB spectrum of the manifolds,
allows also a useful comparison between the FEM and the dual graph method.

We proceed as follows:
\begin{enumerate}
    \item we check the convergence of the method in some cases where the exact LB spectrum on the manifold is known analytically;
    \item we consider two dimensional simplicial manifolds made of regular simplices, 
        where a refinement procedure is available also for the dual graph method, 
    in order to show how both approaches converge to the exact LB spectrum;
\item we notice that our results imply that the standard (not refined) application of both methods yield estimates of the eigenvalues which can be significantly at variance with the ones of the exact LB spectrum, while the refined version (generally unavailable to the dual graph method) shows a good convergence behavior;
\item finally, by means of a toy model, we provide a reason why the eigenvalues obtained with the dual graph method typically undershoot 
    the ones of the exact LB spectrum.
\end{enumerate}

\subsection{Convergence of the method for manifolds with known spectrum}\label{subsec:convtor}

First of all, we check the convergence of FEM to the exact LB spectrum for some manifolds
that can be seen as simplicial manifolds and whose spectrum is known: \textit{flat} toruses.
Indeed, they are nothing but flat parallelepipeds with properly identified boundaries, 
and it is apparent that they can be covered with simplices, though not with regular ones, 
except for some specific ratios of their ``sides''.

For a generic $d$ dimensional smooth torus, with lengths $\{L_{\mu}\}_{\mu=1}^d$, 
the LB spectrum reads:
\begin{equation}\label{eq:dDtorus}
    \mathcal{S}_d \lbrack L_\mu \rbrack \equiv \Big\{ 4\pi^2 \sum\limits_{\mu=1}^{d}\Big(\frac{n_{\mu}}{L_{\mu}}\Big)^2 
    \Big| n_{\mu}\in \mathbb{Z}, \mu=1,\dots,d \Big\},
\end{equation}
where we order the eigenvalues in a non-decreasing fashion, 
and consider degenerate eigenvalues as distinct elements.

\begin{figure}
	\centering
	\includegraphics[width=\linewidth]{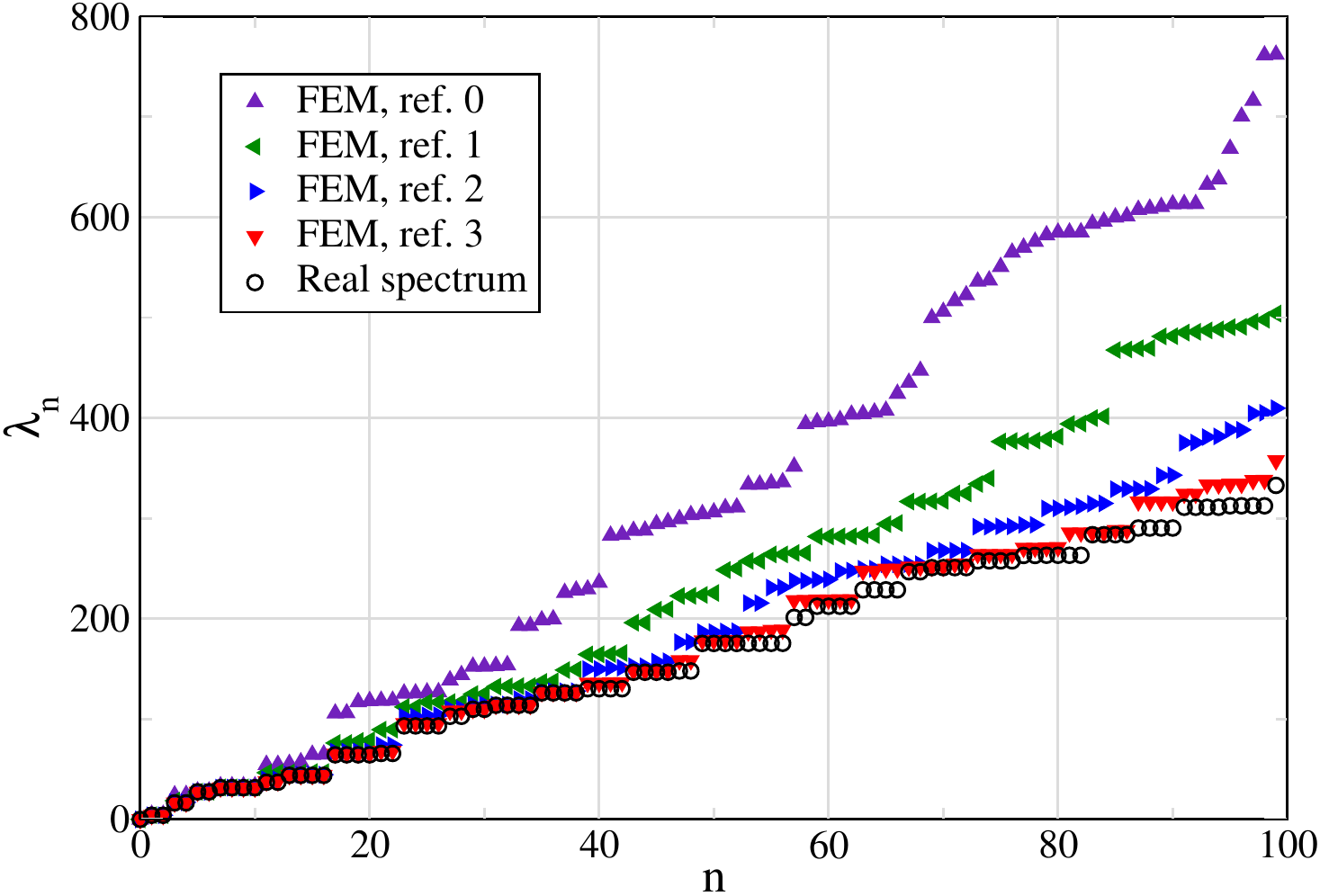}
	\caption{Convergence of the first $100$ eigenvalues, obtained through FEM, 
        to exact spectrum for a 2D flat torus with spatial sizes $L_x=3.1$ and $L_y=1.2$. 
In figure, we show a sufficient number of refinement levels to reach convergence to 
the real spectrum within 1\% in its low part.}
    \label{fig:2dgentorus}
\end{figure}

Figure~\ref{fig:2dgentorus} displays the eigenvalues obtained with FEM at four subsequent 
refinement levels (see Section~\ref{subsec:refinement})
in comparison with the exact LB eigenvalues from Equation~\eqref{eq:dDtorus}, 
for a two-dimensional flat torus with arbitrarily chosen sides $L_x=3.1$ and $L_y=1.2$. 
As expected from a theoretical point of view (see Section~\ref{subsec:fineigp}),
the FEM eigenvalues converge to the exact LB spectrum from above, and are more accurate 
in the the lower part of the spectrum (i.e., for larger scale modes).

\begin{figure}
	\centering
	\includegraphics[width=\linewidth]{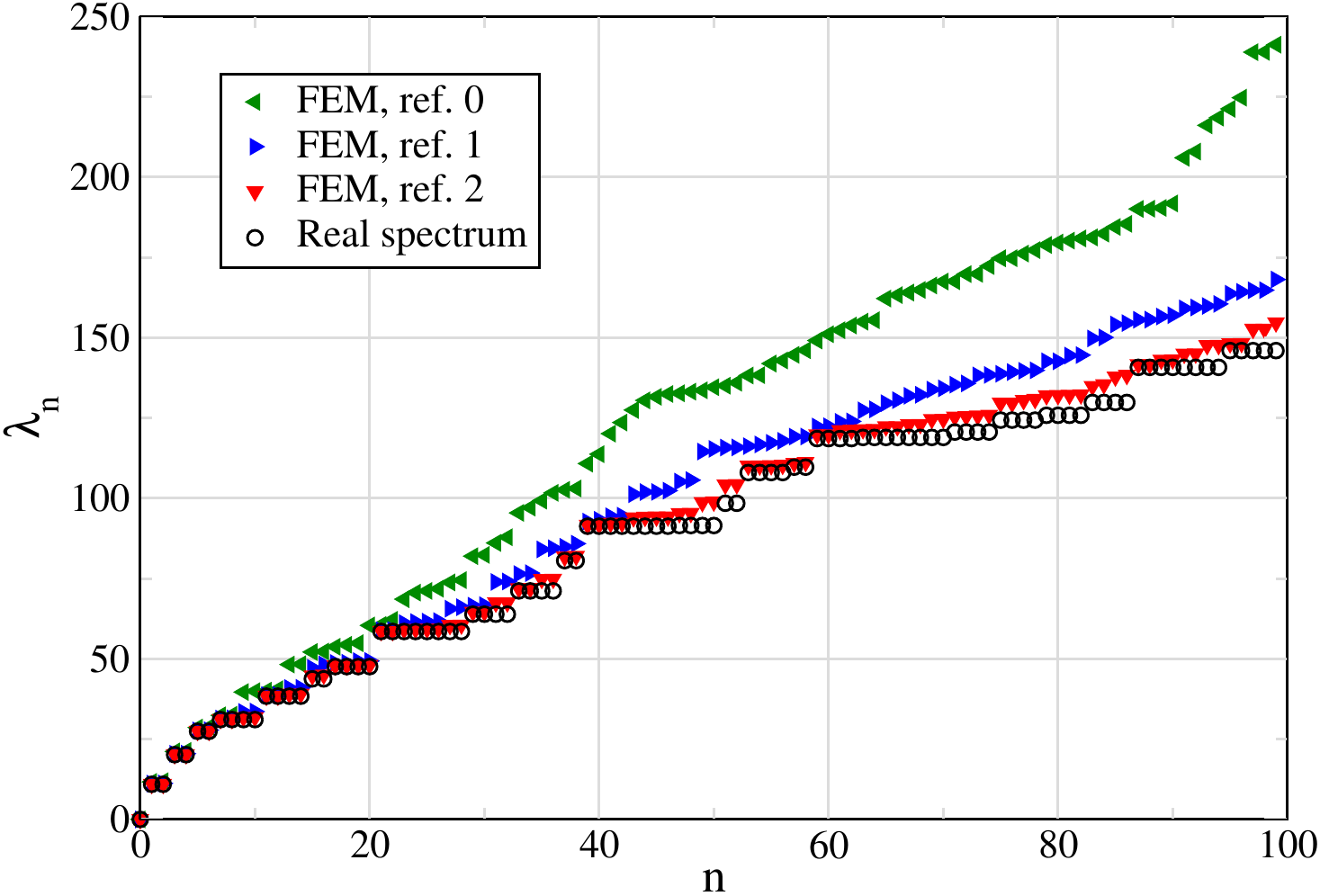}
    \caption{Convergence of the first 100 eigenvalues obtained via FEM to the exact spectrum 
for a 3D flat torus with spatial dimensions $L_x=1.4$, $L_y=1.9$, $L_z=1.2$.}
    \label{fig:3Dtorussp}
\end{figure}

The same behavior is observed for a flat torus in three dimensions;
the only difference found is in the computation time needed to achieve the same precision, 
that becomes greater the higher the dimension, because convergence rate depends on
the \textit{maximal diameter} $h$ among the simplices~\cite{fem_hughesbook,fem_strangbook}, while computation time depends, given our choices, on the number of vertices,
which, for the same decrease in $h$, grows faster in higher dimension. 
Results obtained for a 3D flat torus with $L_x=1.4$, $L_y=1.9$, $L_z=1.2$ 
are shown in Figure~\ref{fig:3Dtorussp}.
A similar behavior is observed also for higher-dimensional toruses, 
which however don't add interesting information to what we have already shown. 

\subsection{Focus on dimension two: refinement for the dual graph method and convergence of FEM for irregular simplicial manifolds}\label{subsec:dim2}

Here we investigate what happens with two-dimensional toruses
which can be covered by equilateral triangles (those with $L_y=\sqrt{3}L_x$), 
as they allow a direct comparison between FEM and the dual graph method. 
Notice that, for such a comparison, the overall numeric factor in Equation~\eqref{eq:dgfinal} 
($4$, in two dimensions) becomes relevant.

\begin{figure}
	\centering
	\includegraphics[width=\linewidth]{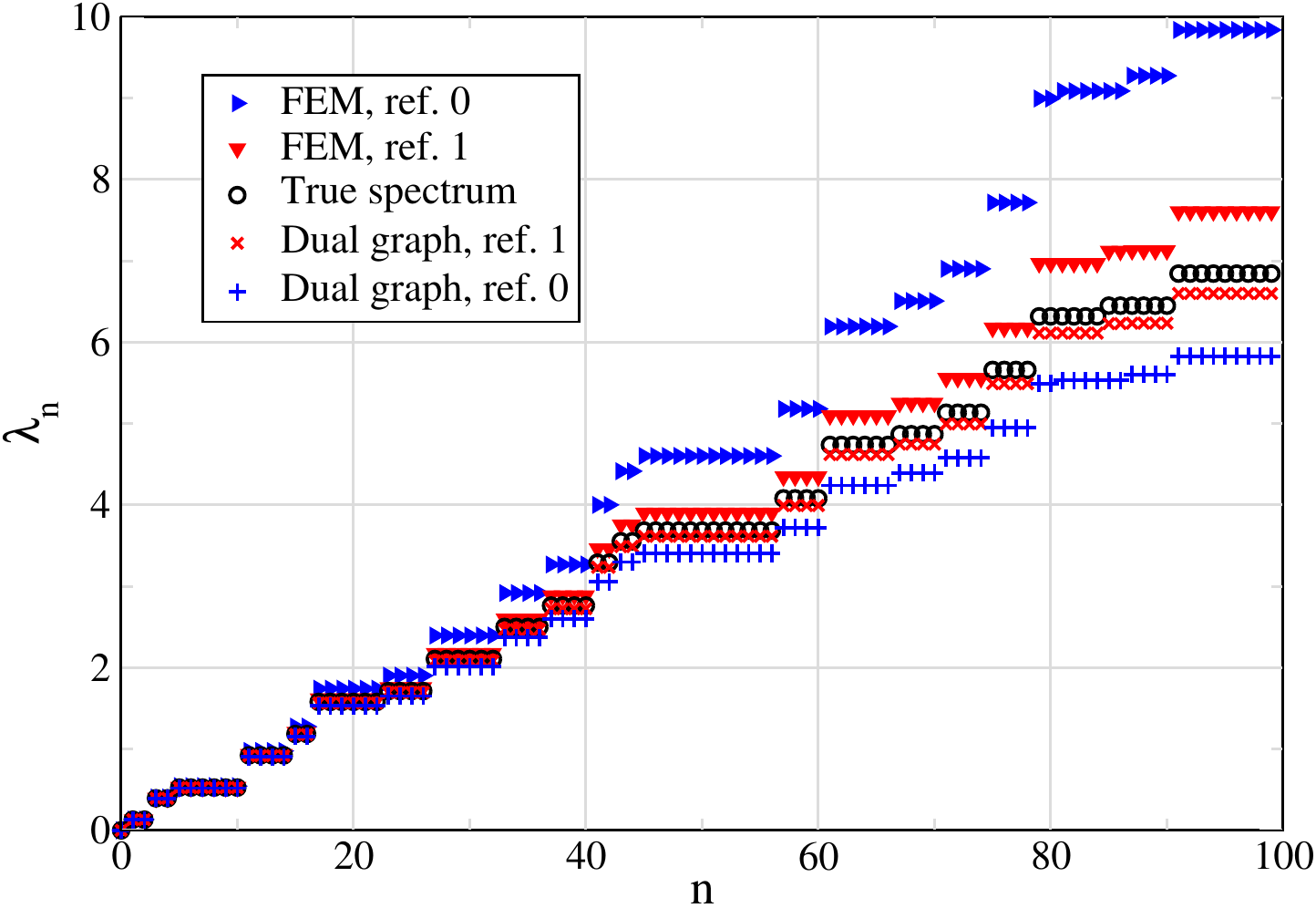}
	\caption{Convergence of the first $100$ eigenvalues, 
    obtained via FEM and dual graph method, for a 2D flat torus made of $20$ rows of $20$ 
    equilateral triangles with unit side, with appropriate identification.
Dual graph eigenvalues have been multiplied by $4$ for comparison with FEM ones, 
as explained in the text.}
    \label{fig:2regtorus}
\end{figure}

As anticipated above, a particularity of dimension two is that we can think of a refinement procedure of the simplicial manifold that allows to iterate the dual graph method: it consists in dividing each equilateral triangle into four new ones by connecting with three new links the middle points of the sides of each triangle.
Figure~\ref{fig:2regtorus} shows the spectra in 
the case of a two-dimensional flat torus made of $20$ rows of $20$ 
equilateral triangles of unit side ($L_x=10$) with properly identified boundaries. 
At increasing refinement level, 
besides the expected convergence of the FEM spectrum to the exact LB spectrum (from above), 
we observe also a convergence of the dual graph spectrum \textit{from below}, 
even if we do not have a well-settled theory ensuring this behavior (unlike in the FEM case).
The interpretation of this behavior will turn out to be fundamental (see Section~\ref{subsec:supr}).

\begin{figure}
	\centering
	\includegraphics[width=\linewidth]{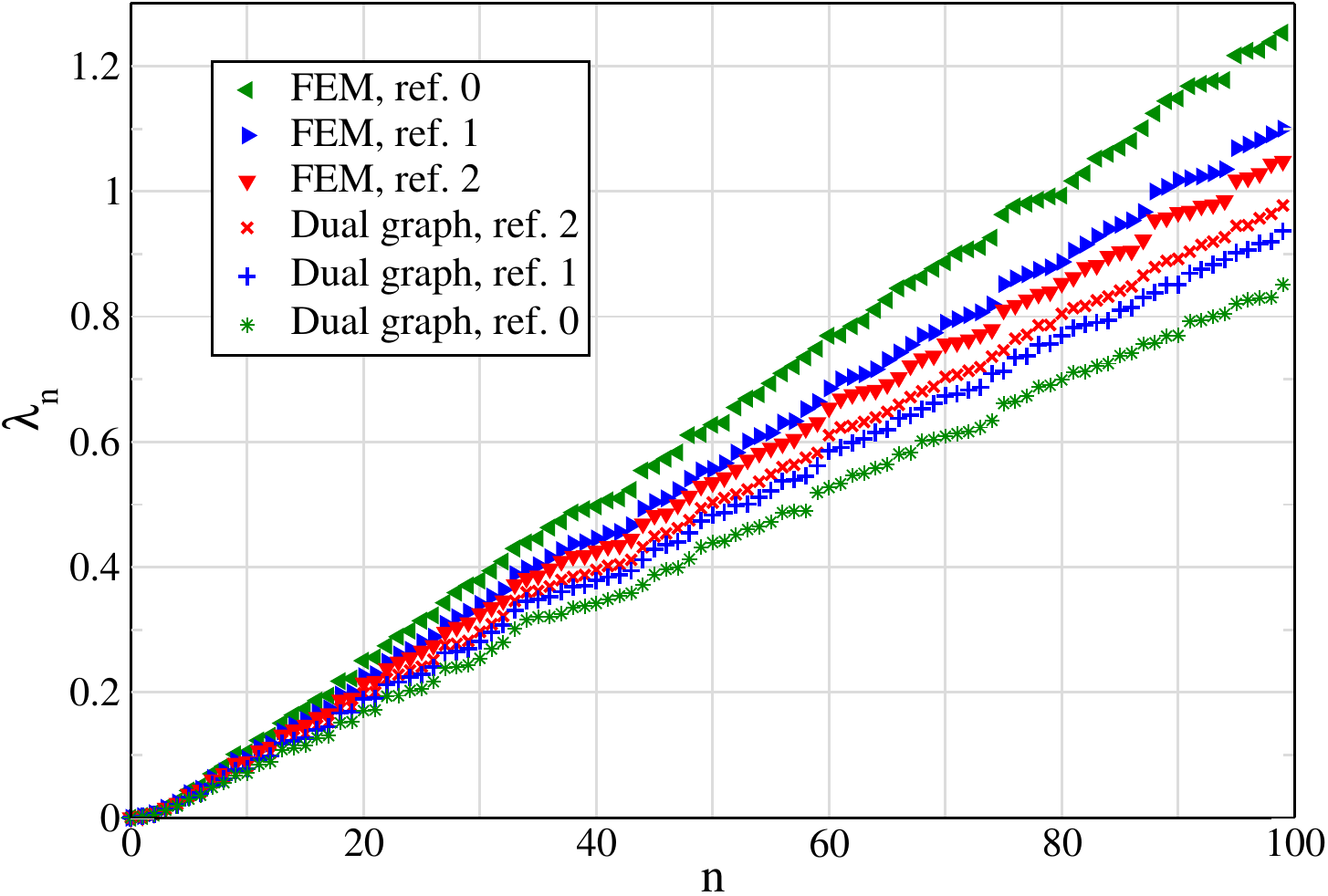}
	\caption{Convergence of the first $100$ eigenvalues of LB operator, 
        discretized by means of the two methods, on a random test CDT configuration 
in two dimensions. The configuration has total volume $2602$ and $1301$ vertices. 
Dual graph eigenvalues have been multiplied by a prefactor $4$ for comparison with FEM ones,
as explained in the text.}
    \label{fig:testconfsp}
\end{figure}

The dual graph method behaves in the same way as in the previous example also on a 
less regular object, like a typical two-dimensional CDT configuration,
whose exact LB spectrum cannot be computed by analytical means.
This helps us check the convergence properties of FEM even on such 
an irregular object: 
in Figure~\ref{fig:testconfsp}, indeed, we can compare the two methods as they 
progressively reach an agreement, following two opposite trends 
(FEM from above and dual graph from below), 
on what can be guessed to be the exact LB spectrum on this real-life CDT object. 
Not very surprisingly, computation times are worse than for more regular geometries.

\subsection{General inaccuracy of the dual graph method in the case of no refinement steps}\label{subsec:inacc}

\begin{figure}
	\centering
	\includegraphics[width=\linewidth]{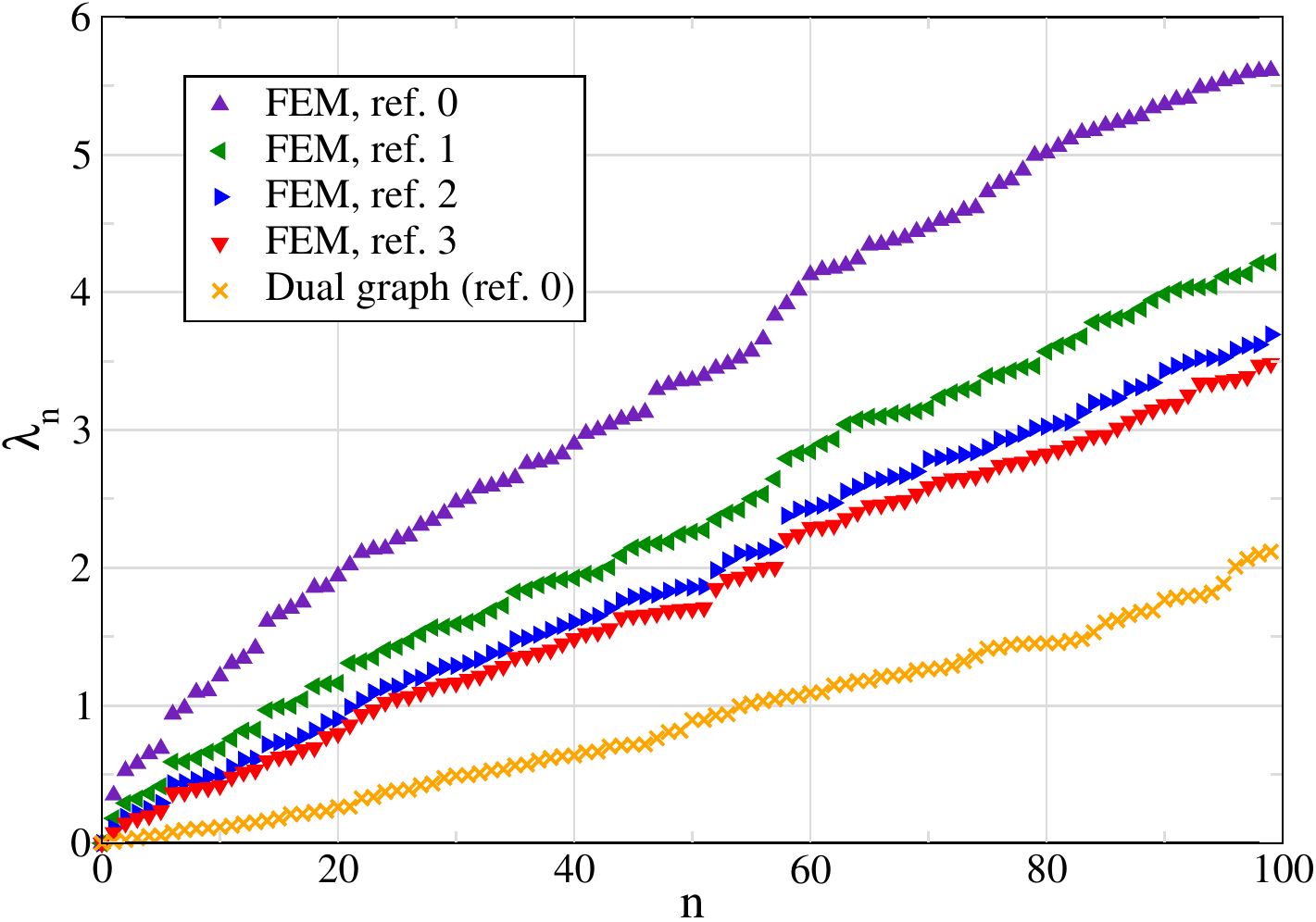}
	\caption{Convergence of the first $100$ eigenvalues of LB operator, 
    obtained by means of the two methods, 
on a spatial slice (with $V_S=2631$) of a random test CDT configuration in four dimensions. 
Dual graph eigenvalues have been multiplied by $9$ for comparison with FEM ones,
as explained in the text.}
    \label{fig:exslice}
\end{figure}

Besides being a check for the convergence of FEM, 
the two examples of the previous section 
testify the fact that the spectrum of the Laplace matrix of the 
original \emph{unrefined} dual graph
quantitatively differs in a non-negligible way from the exact LB spectrum: 
indeed, we stress again that, in more than two dimensions, 
no refinement for the dual graph is available, 
so the situation one faces is the same as in Figure~\ref{fig:testconfsp} 
but with only the $0$th refinement (the original triangulation). 
An example of this is shown in Figure~\ref{fig:exslice}, 
depicting the spectrum obtained with the two methods on a spatial slice
of a 4D CDT configuration coming from a point in $C_{dS}$ phase 
(using $9$ as the overall factor for the dual graph method). 

What we show in the next section is that the difference of the dual graph spectrum 
from the exact one can be so large that it could lead to misrepresent even some important (large scale) \textit{qualitative} features of the simplicial manifold. 

\subsection{Issues of the dual graph method: discussion and a toy model}\label{subsec:supr}

In order 
to understand the reasons for the quantitative discrepancies shown in the previous section between 
the (unrefined) dual graph spectrum and the exact LB spectrum, 
it is important to establish why the first seems to systematically underestimate the second, 
as apparent  from Figures~\ref{fig:2regtorus},~\ref{fig:testconfsp} and~\ref{fig:exslice}.

Regarding this observation, we have a tentative explanation, 
which we think may prove to be quite compelling after we show 
a simple toy model serving as a worst-case scenario.

\begin{figure}
	\centering
	\includegraphics[width=\linewidth,trim={3.7cm 1.8cm 2.7cm 2.05cm},clip]{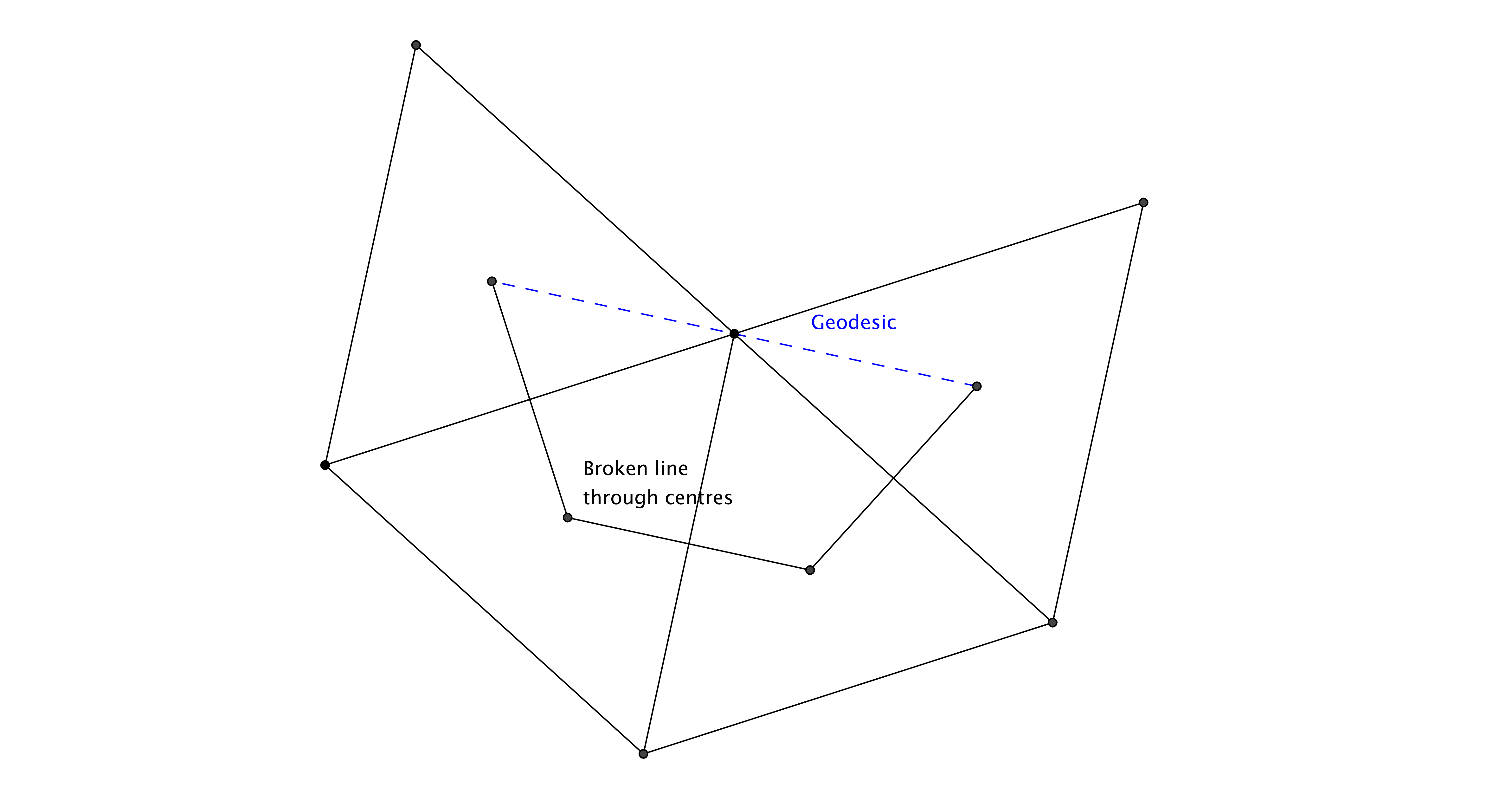}
	\caption{An example of the physical distance between two centers 
    being overestimated by the path through centers connecting them.}
    \label{fig:shortcut}
\end{figure}

As explained in Section~\ref{subsec:difft}, if we consider a diffusion process 
on a manifold, the eigenvalues of the LB operators 
are associated with the typical time rates of the diffusion modes and to 
the typical length scales of the manifold. 
When we build the dual graph associated with a simplicial manifold, 
we lose part of the metrical information that is relevant to a diffusion process 
on that object, 
which is not merely made of the adjacency relations between simplices.
A diffusion process on a simplicial manifold takes place in 
\textit{the whole physical space} of which it is constituted,
and not only along the segments connecting the centers of adjacent simplices: 
the main consequence of this fact is that in the dual graph case 
the \textit{distance} between two centers ``perceived'' by a diffusing particle 
is incorrectly represented by the adjacency matrix, 
as if the particle was constrained to diffuse along the segments joining the centers 
instead of going from one point to another along the shortest path 
(i.e., a geodesic of the simplicial manifold). 
This happens because we have lost memory of the metric al information 
(and, in some sense, also of the topology, as we will show in the toy model) 
of everything that is not the centers, 
the edges between them and the angles between the edges 
(which the correspondence to the LB operator relies upon). 
Therefore, we claim that all typical length scales are overestimated and the error 
depends on how badly the geodesics are approximated by broken lines passing through 
the centers, thus resulting in the underestimation of 
the eigenvalues they are associated with; 
this idea is depicted in Figure~\ref{fig:shortcut}.

In light of these considerations, 
the observed convergence from below of the eigenvalues in two dimensions, 
where a meaningful refinement procedure is available, 
can be naturally understood
as the fact that shortest paths connecting centers in the dual graphs 
approach more and more, at increasing refinement level, 
the true geodesics of the simplicial manifold.

The information we are losing by neglecting 
the flat interior of simplices does not have to be taken into account
when we set up simulations using the Einstein-Hilbert action 
in the coordinate-free Regge formalism, 
because that requires only the total volume and the total curvature, 
information that can be represented in terms of combinatorial observables.
However, in order to describe the geometry of a simplicial manifold 
in terms of local observables (like the propagation or diffusion of test fields) 
one has to take into consideration the whole geometrical structure, including that information.

An objection to the necessity of the FEM representation 
as a substitute to the dual graphs one could be that, 
since the interior of a simplex in the original triangulation is flat, 
and its sizes are comparable with the lattice spacing, 
the results expected using dual graph techniques 
(including non-spectral ones) would have no substantial 
effect on large scale observables, whose correlation lengths 
are assumed to be much larger than the lattice spacing, 
therefore making dual graph results inaccurate just only for small 
scales features which would be discarded anyway.
However, even if this argument may work sometimes 
(e.g., for hypercubic lattices representing 
flat spaces), it doesn't hold in this case.
The key fact is that the length scales provided by the dual graph method do not correspond to actual physical scales of the manifold in a clear way, and the relation depends on the geometric properties of the manifold under analysis, as
the geodesic overestimation resulting from the dual graph representation 
can severely and differently impact the observables, even at large scales.
This geodesics overestimation can be responsible for \textit{arbitrarily} poor estimates
when the total volume goes to infinity\footnote{A situation we have to deal 
    with in the application to CDT, 
where it represents the infinite volume limit}, 
being it possible to make the dual graph method detect the biggest length scale 
as going to infinity while it is actually staying finite. 
This would result in the dual graph method yielding a vanishing spectral gap 
in the infinite volume limit, with the real LB spectrum actually having a lasting non-zero gap.

Indeed, it is not hard to realize that the worst case of distance overestimation
happens in the neighborhood of $(d-2)$-simplices 
with high coordination numbers (i.e., those associated with a high local negative curvature): 
the geodesic distance between the centers of two ``opposite'' $d$-simplices 
sharing a $(d-2)$-simplex with many $d$-simplices surrounding it, 
that is roughly the diameter of the $d$-simplices, 
is very badly estimated by a broken line passing through the centers, 
that is a very long (half) loop around the $(d-2)$-simplex. 
This observation inspires the construction of a simple 2D model that we will now discuss:
we consider an arbitrary number of triangles all sharing the same vertex, 
with each of the links coordinated to it in common between two of them; 
then, in order to make the manifold boundaryless, 
we take a second identical ``sheet'' of triangles forcing each triangle of 
the first sheet to share the third side with
the third side of the corresponding triangle of the second sheet. 
In this setting, we show how by increasing the total volume the FEM correctly 
detects an almost constant non-vanishing spectral gap, 
while the dual graph method yields a vanishing one, 
since it represents this object essentially as a discretization of $S^1 \times \{0,1\}$.

\begin{figure}
	\centering
	\includegraphics[width=\linewidth]{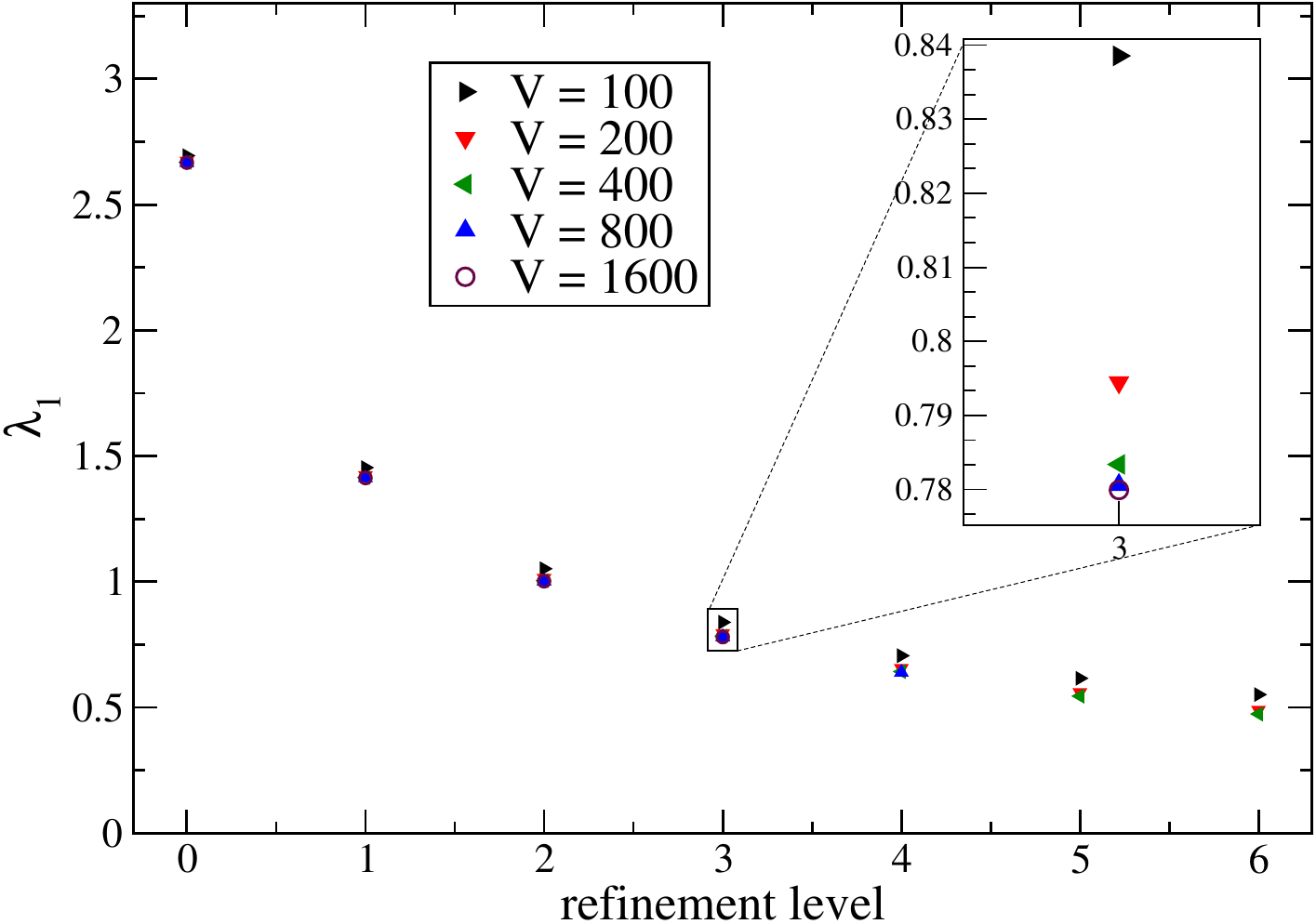}
	\caption{Estimate for the spectral gap as a function of the refinement step 
    for the toy model at  various volumes. For a given refinement step, 
the estimates for the spectral gap are very similar for big volumes, 
thus the exact value can be expected to be essentially the same too.}
    \label{fig:toygap}
\end{figure}

Because of the highly pathological geometry, 
it is not easy to push the FEM to convergence, 
that in this particular case we see to be slower for lower-order 
eigenvalues than for higher-order ones: 
within reasonable computation times we could achieve stable estimates (within 1\%) 
of the orders between, say, $n=10$ and $n=20$, but not of the spectral gap. 
Given that we have a specific interpretation of the meaning of the spectral gap, 
we prefer, anyway, to focus on it rather than on other orders: 
even without explicit achievement of convergence, indeed, 
in Figure~\ref{fig:toygap} we can see how when the volume increases 
(we use as progression $V=100,200,400,800,1600$) the spectral gap seen at 
each refinement step by the FEM is almost constant, with values that become 
more and more similar at increasing volume as can be seen in the inset plot.
The exact LB spectral gap, i.e., the one at infinite refinement level, 
gets the same non-zero value independently on the volume. 
This is expected, the diffusion rate being more related on the finite manifold diameter 
(the maximum of minima among path lengths) than on its volume.

\begin{figure}
	\centering
	\includegraphics[width=\linewidth,trim={0.45cm 0.30cm 1.5cm 1.35cm},clip]{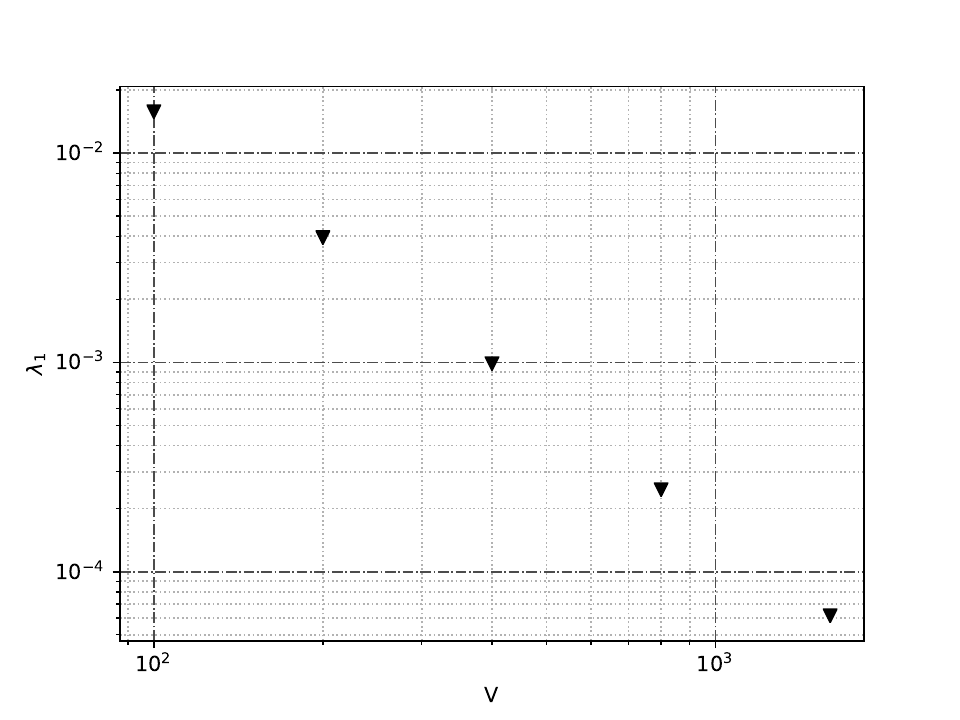}
	\caption{Spectral gap detected via dual graph method 
        for the toy model at various volumes. The value vanishes quadratically as expected. 
    Both scales are logarithmic.}
    \label{fig:toygraph}
\end{figure}

The dual graph method, instead, detects a spectral gap quadratically decaying 
to zero for the same volume progression as above, 
as one can predict by considering the aforementioned $S^1$ structure of 
the dual graph of this simplicial manifold. This is shown in Figure~\ref{fig:toygraph}.

We have seen, then, that the dual graph method gives estimates that not only 
quantitatively differ from the exact LB spectrum on the manifold 
(with the consequent different estimates for related spectral observables, as we show below) 
but can also fail to reveal some important qualitative geometrical features 
as a finite diameter for infinite volume, a situation that might even result 
not to be irrelevant for CDT, where infinite volume represents the thermodynamic limit, 
even if the randomness of geometry would probably prevent such extreme situations from happening.

\section{Numerical results}\label{sec:numres}
Having explained our motivations for applying 
FEM to the study of the LB spectrum on simplicial manifolds, 
here we show it in action on CDT, where we obtain results that   
significantly differ from analogous ones obtained by using the dual graph method. 
In particular, for the sake of comparison, 
we considered the large-scale dimension of spatial slices in 
$C_{dS}$ phase and the critical index of the $C_b$-$C_{dS}$ transition 
along the line $k_0=0.75$, which were already investigated in~\cite{lbrunning} 
using dual graph methods.

Both of these results involve the spectral analysis of spatial slice submanifolds, 
for which the thermodynamic limit is assumed to coincide with the behavior 
at large spatial volume $V_S$: 
the underlying assumption is that the overall volume fixing, 
if large enough (in all of our simulations $V_{S,tot}=80k$), 
does not significantly affect the features of spatial slices. 
This assumption should be verified by checking the stability of 
the results under the increase of the total spatial volume, 
a computationally demanding operation that we postpone to future works, 
our main aim now being to show the potential of these methods. 
Furthermore, since computational costs for even the low part of the FEM spectrum
grow rapidly when the refinement level increases,
we need to resort to an extrapolation at infinite refinement level, as we discuss next.\\
                           
Summarizing, the following three limiting procedures should be performed in this specific order:
\begin{itemize}
    \item for each simplicial manifold $\mathcal{M}$, extrapolation of the individual FEM eigenvalues to ``infinite refinement level'' $\lambda_n^{(r)}[\mathcal{M}] \xrightarrow{r\to \infty} \lambda_n^{(\infty)}[\mathcal{M}]$, in order to obtain an accurate enough approximation of the spectrum of the exact LB differential operator on $\mathcal{M}$;
    \item for each ensemble of configurations at specific values of the parameters, 
        one should first perform an average of the eigenvalues at each specific order $n$ 
        and then take the thermodynamic limit (i.e., infinite volumes in lattice units) 
        \mbox{${\langle \lambda_n \rangle}_V \equiv \frac{1}{\lvert \mathcal{C}_{V} \rvert} \sum_{\mathcal{M}\in \mathcal{C}_V} \lambda_n^{(\infty)}[\mathcal{M}] \xrightarrow{V\to \infty}  {\langle \lambda_n \rangle}_\infty$} by considering the spectra of ensembles
        $\mathcal{C}_{V}$ with increasing volumes;
    \item study the critical scaling of the twice-extrapolated eigenvalues ${\langle \lambda_n \rangle}_\infty {(k_0,\Delta)}$ observed as the phase transition is approached.
\end{itemize}

\subsection{Extrapolation to infinite refinement}

Both the results we are going to show are relative to the LB spectrum on the spatial slices, 
so we have to deal with great variability in the volumes:
first of all, then, we group the slices in a proper number of volume bins 
(excluding too small ones). 
This is anyway necessary, regardless of the extrapolation step\footnote{Since 
    the systematical error of the extrapolation for each order $n$ and on 
each single slice is expected to be smaller than the statistical error due to the variability 
of the eigenvalue estimate on a volume bin, it is useful to proceed by extrapolation 
only after volume binning instead of doing as described ideally 
in the scheme oulined in the list above.}, 
in order to study the thermodynamic limit; indeed, this was done also when
the dual graph method was used (see~\cite{lbrunning}).
Then, for each fixed refinement level, 
we average the eigenvalues of each fixed order $n$ of the slices in each volume bin, using the standard deviation of the mean as a measure of uncertainty;
finally, we extrapolate the value for \textit{infinite refinement level} 
for each order and bin, that we read as a reasonable estimate of the average 
in that volume bin of the exact LB eigenvalues of that order.

The functional form we use for our extrapolation is
\begin{equation}\label{eq:extrapolation}
\lambda_n^{(r)} = \lambda_n^{(\infty)} + A_n e^{-r/B_n},
\end{equation}
where r is the refinement level, $A_n$ and $B_n$ two parameters that depend on 
the order of the eigenvalue and on the volume bin 
and $\lambda_n^{(\infty)}$ the extrapolated value. 
The reason why we use this form is that the most relevant parameter for the 
convergence of FEM is the \textit{maximal diameter} $h$ among the simplices 
of the triangulation: 
the convergence to the exact LB eigenvalues~\cite{fem_hughesbook,fem_strangbook}, 
given that the simplices are not too pathological (e.g., having very acute angles), 
can actually be faster than a power of $h$.
Assuming our ``average'' convergence to happen \textit{exactly} according 
to a power of $h$, from the fact that in our case $h$ is halved 
at each refinement step, we obtain:
\begin{equation}
\lambda_n^{(r)} = \lambda_n^{(\infty)} + h^k = \lambda_n^{(\infty)} + h_0 2^{-rk},
\end{equation}
that can be rewritten in the form of Equation~\eqref{eq:extrapolation}.
We show below that our data are in good agreement with this picture.

\subsection{Large-scale spectral dimension of spatial slices in $C_{dS}$ phase}

In previous studies with dual graphs~\cite{lbstruct,lbrunning,lbpos19}, 
the large-scale effective dimension of the spatial slices in phase $C_{dS}$ seemed
to be almost independent of the point in the $C_{dS}$ phase, 
and compatible regardless of the definition of dimension: 
the ``diffusive'' one and the one based on ``energy-levels'', 
both introduced in Section~\ref{sec:LBspectrum_geom}.

It was reasonable to choose to analyze two of the phase space points in which 
the known estimates were obtained, that is, 
$(k_0,\Delta)=(2.2,0.6)$ and $(k_0,\Delta)=(0.75,0.7)$. 
The maximum number of refinements levels we could analyze 
with the resources available was $\bar{r}=3$ 
(besides the starting triangulation, labeled as $r=0$). 
In both cases, the bins we used were of equal volume extent and were chosen 
in such a way that the slices were, more or less, equally distributed; 
moreover, 
we excluded from the analysis slices with volumes $V_S < 500$, 
for which finite volume effects could be significant.
The best-fit of the extrapolations performed for each volume bin, 
according to the functional form in 
Equation~\eqref{eq:extrapolation}, shows, in almost every case, a good agreement 
between data and our chosen heuristic form, 
with $\chi^2$ ranging between $0.5$ and $3.5$ (one degree of freedom). 
Then, for both of the two points we performed two different best-fit procedures 
to find an estimate of the large-scale spectral dimension. 
The first form considered was:
\begin{equation}\label{eq:cdsphasefit}
\langle \lambda_n \rangle = A_n V_S^{-2/d_{EFF}},
\end{equation}
for a global fit using the first ten eigenvalue orders with $d_{EFF}$ and $\{A_n\}_{n=1}^{10}$ 
used as free parameters ($d_{EFF}$ in common for every order $n$).\\ 
The second form considered was:
\begin{equation}\label{eq:cdsphasefit2}
\langle \lambda_n \rangle = (n/V_S)^{2/d_{EFF}},
\end{equation}
that, as the previous one,
can be obtained from Equation~\eqref{eq:elevspdim}, 
again for the first ten orders, 
where the effective dimension $d_{EFF}$ is the only free parameter 
and corresponds to the dimension of the simplicial manifold 
at the largest scale (lowest part of the spectrum).

At each point, we also performed the same two procedures by using the values that 
could be extrapolated from the first two refinements only, 
in order to have a further indication of the goodness of our heuristic extrapolation 
method by confronting the obtained results with ($\bar{r}=3$) 
and without ($\bar{r}=2$) using the third refinement. 
We saw general compatibility of the extrapolated values of $\langle \lambda_n \rangle$, 
with a slight systematic overestimation in the case 
without refinement $3$ with respect to the other.

\begin{figure}[ht!]
	\centering
	\includegraphics[width=\linewidth,trim={0.7cm 0.3cm 0.7cm 1.35cm},clip]{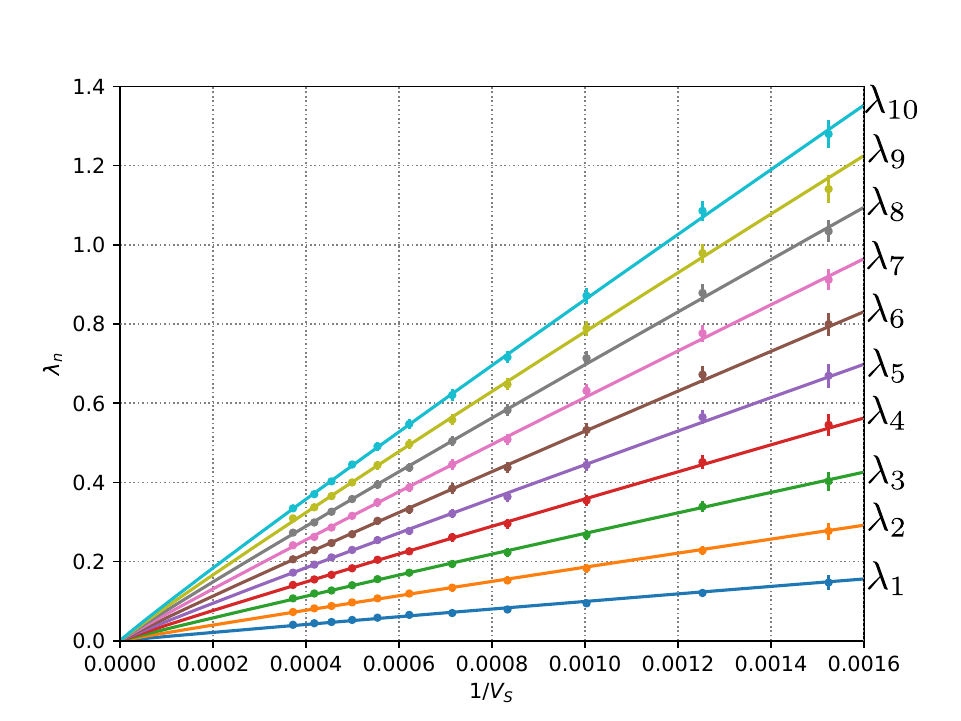}
	\caption{Extrapolated first ten orders of eigenvalues vs $1/V_S$ for 
    big enough slices ($V_S>500$), with best-fit curves according to 
Equation~\eqref{eq:cdsphasefit} (common $d_{EFF}$). 
Phase space point $k_0=2.2, \Delta=0.6$, volume fixing $V_{S,tot}=80k$.}
    \label{fig:2206}
\end{figure}

\begin{figure}[ht!]
	\centering
	\includegraphics[width=\linewidth,trim={0.72cm 0.3cm 1.1cm 1.35cm},clip]{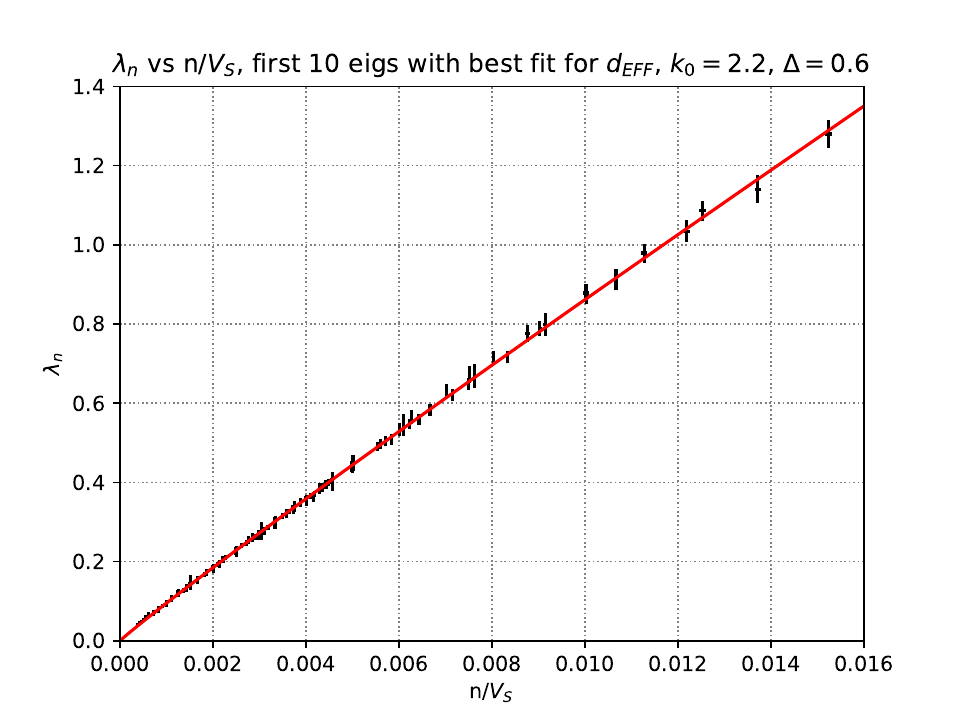}
    \caption{Same as in Figure~\ref{fig:2206}, but using Equation~\eqref{eq:cdsphasefit2}
        as fitting function, and data collapsed by using $n/V_S$ as independent variable.}
    \label{fig:2206agg}
\end{figure}

Fit results for the point $(k_0,\Delta)=(2.2,0.6)$ are shown in Table~\ref{tab:fit2206}, 
where the functional forms in Equations~\eqref{eq:cdsphasefit} and~\eqref{eq:cdsphasefit2} 
and with refinements up to $\bar{r}=2$ and $\bar{r}=3$ have been considered independently.
Figure~\ref{fig:2206} represents extrapolated eigenvalues and best-fit with the function in 
Equation~\eqref{eq:cdsphasefit} including the third refinement ($\bar{r}=3$), 
while Figure~\ref{fig:2206agg} shows the same 
in the case of a fit with the function in Equation~\eqref{eq:cdsphasefit2}.
The compatibility between the estimates with ($\bar{r}=3$) 
and without ($\bar{r}=2$) using the third refinement is not exceptional but neither terrible, 
and, in general, the first estimate should be preferred, of course, 
as it involves a broader dataset.


\begin{table}
\begin{tabular}{l|l|l}
\hline
Fit function  & $d_{EFF}$ [$\chi^2/dof$] Eq.~\eqref{eq:cdsphasefit} & $d_{EFF}$ [$\chi^2/dof$] Eq.~\eqref{eq:cdsphasefit2} \\ \hline
$\bar{r}=2$ & $2.129(34)$ [$4/99$] &  $2.14(3)$ [$5/108$] \\ \hline
$\bar{r}=3$ & $2.084(16)$ [$15/99$] & $2.091(12)$ [$16/108$]  \\ \hline
\end{tabular}
\caption{Fit results of the functional forms 
in Equations~\eqref{eq:cdsphasefit} and~\eqref{eq:cdsphasefit2} 
for the FEM extrapolations of the first ten orders of eigenvalues 
(see Figures~\ref{fig:2206} and~\ref{fig:2206agg})
for spatial slices of configurations at the point $(k_0,\Delta)=(2.2,0.6)$.}
\label{tab:fit2206}
\end{table}

In general, as it can be seen, data well fit a large-scale finite and 
fixed effective dimension, and the two different best-fit procedures 
return compatible estimates for the effective dimension, 
but this dimension significantly differs from the previously 
known value of about $1.6$ found in~\cite{lbstruct}. 
We choose as our conservative estimate the average of the two most reliable 
estimates ($\bar{r}=3$): $d_{EFF}=2.088(18)$ (with the semidispersion as error).

As for the point $(k_0,\Delta)=(0.75,0.7)$,
we performed the same kind of analysis, obtaining the results shown in Table~\ref{tab:fit07507}.
The plots for this point of the phase diagram are qualitatively similar 
to~\ref{fig:2206} and~\ref{fig:2206agg}, and will not be shown.
As before, the results are somewhat compatible, 
and the estimates with $\bar{r}=3$ are preferred.
Again, we see the expected fixed and finite-dimensional behavior, 
and the estimated dimension significantly differs from that 
of about $1.6$ found in~\cite{lbrunning}: our conservative estimate, 
indeed is $d_{EFF}=2.202(16)$.

\begin{table}
\begin{tabular}{l|l|l}
\hline
Fit function  & $d_{EFF}$ [$\chi^2/dof$] Eq.~\eqref{eq:cdsphasefit} & $d_{EFF}$ [$\chi^2/dof$] Eq.~\eqref{eq:cdsphasefit2} \\ \hline
$\bar{r}=2$ & $2.28(3)$ [$7/99$] &  $2.25(2)$ [$10/108$] \\ \hline
$\bar{r}=3$ & $2.216(15)$ [$23/99$] & $2.187(11)$ [$34/108$]  \\ \hline
\end{tabular}
\caption{Fit results of the functional forms 
in Equations~\eqref{eq:cdsphasefit} and~\eqref{eq:cdsphasefit2} 
for the extrapolations of the first ten orders of eigenvalues 
for spatial slices of configurations at the point $(k_0,\Delta)=(0.75,0.7)$.}
\label{tab:fit07507}
\end{table}

Interestingly, the two final estimates, besides being incompatible with the ones previously 
known in literature, appear to be quite different for the two phase space points, 
while up to now it has been believed that this dimension is almost constant across 
the $C_{dS}$ phase. This may be worthy of further inquiry as it might turn out 
to be significant, for example, for the study of the RG flow properties in that phase.

\subsection{Critical index of $C_b$-$C_{dS}$ transition}

The most interesting result from~\cite{lbrunning} 
is the analysis of the critical behavior of the low spectrum
of $B$-type slices in phase $C_b$ while approaching 
the $C_b$-$C_{dS}$ transition along two lines of constant $k_0$. 
Here we compare the results from~\cite{lbrunning}, obtained using dual graph methods, 
with the application of FEM to the same configurations, in particular, 
checking the value of the critical index $\nu$ in the shifted power law:
\begin{equation}\label{eq:critscaling}
\langle \lambda_n \rangle_{\infty} = A_n (\Delta_{crit} - \Delta)^{2\nu}.
\end{equation}
We analyzed some points along the line $k_0=0.75$ only, 
and considered the critical scaling of the first ten orders of eigenvalues at the same time. 
First, we had to extrapolate to infinite refinement level, 
which we did in the same way as described above using Equation~\eqref{eq:extrapolation} 
on data coming from refinements $0$ up to $\bar{r}=3$, 
and then to the thermodynamic limit ($V_S \to \infty$) for each order.
For this second procedure, not having precise expectations on the large-scale behavior 
but the fact that each order should not approach zero, 
we followed~\cite{lbrunning} and used the simplest form compatible with data, 
that is, a quadratic polynomial in $1/V_S$:
\begin{equation}\label{eq:thermlim}
\langle \lambda_n \rangle = \langle \lambda_n \rangle_{\infty} + \frac{A_n}{V_S} + \frac{B_n}{V_S^2}.
\end{equation}

\begin{figure}
	\centering
	\includegraphics[width=\linewidth,trim={0.72cm 0.27cm 0.70cm 1.38cm},clip]{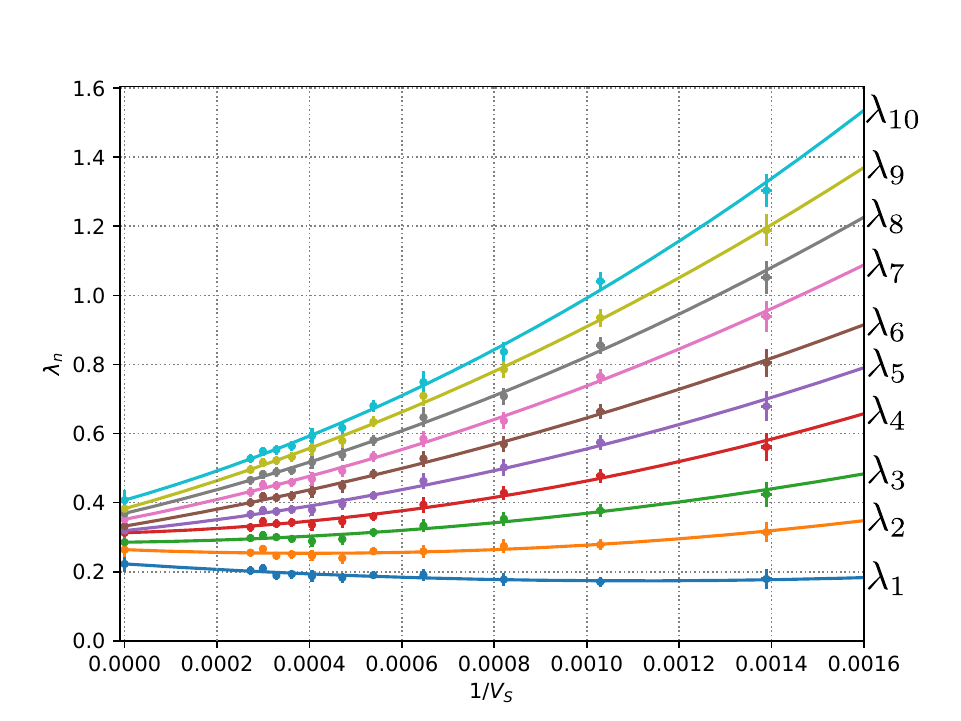}
	\caption{First ten eigenvalue orders of spatial slices (with $V_S>500$) 
    vs $1/V_S$, with extrapolation to the thermodynamic limit ($V_S\rightarrow\infty$) 
in $C_b$ phase. Phase space point $(k_0,\Delta)=(0.75,0.7)$, total spatial volume $V_{S,tot}=80k$.}
    \label{fig:thermlim}
\end{figure}
For every phase space point taken into account, both procedures 
gave satisfactory results in terms of the agreement between data and model: 
for the infinite refinement extrapolations, the chi-squared was $2$ in the worst case 
(with $1$ dof), 
while the thermodynamic limit extrapolations always yielded $\chi^2/dof<1$. 
For illustration purposes, Figure~\ref{fig:thermlim} displays the extrapolation 
to the thermodynamic limit for the first ten eigenvalue orders in the phase space point 
$k_0=0.75, \Delta=0.575$.
\begin{figure}[h]
	\centering
	\includegraphics[width=\linewidth,trim={0.72cm 0.36cm 1.13cm 1.45cm},clip]{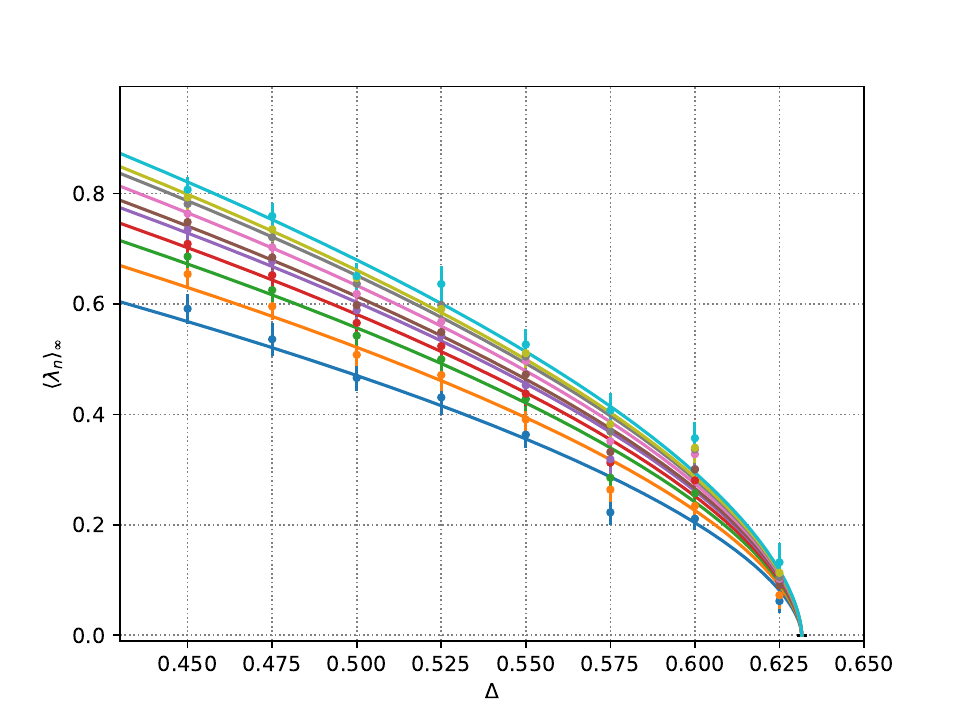}
	\caption{Critical behavior of the first ten eigenvalue orders along the 
    line at fixed $k_0=0.75$ and varying $\Delta$, with best-fit curves of the form shown 
in Equation~\eqref{eq:critscaling} with common $\nu$ and $\Delta_{crit}$.
    Curves of increasing eigenvalue order are shown from below to above in the plot.}
    \label{fig:critscaling}
\end{figure}
We then analyzed the first ten eigenvalue orders by fitting our data with 
Equation~\eqref{eq:critscaling}, forcing the critical index $\nu$ and the 
critical point $\Delta_{crit}$ to be the same for every order. 
We used data coming from eight phase space points with $\Delta$ ranging 
from $0.45$ to $0.625$: we chose not to go too deep inside $C_b$ phase because of 
the influence of the expected sub-dominant terms of the critical scaling 
and excluded them by checking the stability of our estimate of $\nu$ under 
the removal of the points with lower $\Delta$ parameter. 
We obtained, as best-fit parameters $\nu=0.293(10)$ 
and $\Delta_{crit}=0.6316(15)$, with $\chi^2 \approx 67$ ($68$ dof); 
our data and best-fit curves are displayed in Figure~\ref{fig:critscaling}.

It is apparent that data (from a bigger dataset) are still compatible with 
the critical scaling found in~\cite{lbrunning}, 
but, while the estimated location of the transition line agrees with 
the previous findings, the value we found for the critical index \textit{significantly differs}
from the previous estimate. We remark that a difference like this may be 
of great importance if critical indexes of different observables have 
to be compared to find a physical continuum limit in the phase diagram.

\section{Summary and Conclusions}\label{sec:conclusions}

In this work, we have reviewed some concepts about spectral analysis 
on simplicial manifolds using the dual graph representation, 
discussing its domain of definition (triangulations with equilateral simplices, 
which can be mapped to undirected graphs), and its limitations, it
being an approximation to the Laplace-Beltrami operator (in the sense recalled in Section~\ref{subsec:lap_dg}) acting on the whole infinite-dimensional $H^1$ space on the simplicial manifold.

In order to extend the use of the spectral observables beyond this domain, we introduced the Finite Element Method formalism and its application to the solution of the LB eigenproblem (Section~\ref{sec:femintro}, with details left to Appendix~\ref{sec:details}). The two representations, dual graph and FEM,
were compared on a series of test geometries in Section~\ref{sec:examples},
showing that, while the dual graph method and FEM display convergence to the same spectrum
when a refinement procedure is applicable in both cases
(that is, only for two-dimensional equilateral triangulations), 
when this procedure is not available, the results provided by the two methods differ in a non-negligible way.

We tried to explain this disagreement between the dual graph and FEM representation by identifying the main reasons in the phenomenon of geodesics overestimation, and in the lack of a general procedure ensuring convergence to the exact spectrum of \emph{the LB differential operator}(which is linked to the geodesics of the simplicial manifold in the way outlined in Section~\ref{subsec:difft}); this spectrum is instead obtained as the limit of the FEM procedure at infinite refinement level, ``curing'' the geodesic overestimation. We showed that this overestimation can affect also the large scale behavior of observables, in particular when large loops in the dual graph are involved (a loop being made of many $d$-simplices encircling around a $d-2$-simplex), as discussed using an illustrative toy model in Section~\ref{subsec:supr}.

Finally, in Section~\ref{sec:numres}, we compared some earlier dual graph 
spectral results on CDT spatial slices with the results of the application of the FEM approach.
We have shown that the FEM effective dimension observed on spatial slices in two points 
of the phase diagram in the $C_{dS}$ phase region is significantly different from
the one observed using dual graph techniques, 
and that while the dimensions for the two points 
seemed to be compatible in the dual graph case, with $d_{EFF}\simeq 1.6$, 
they are detected as incompatible in the FEM case.
We have also investigated again, but using FEM techniques, 
the critical behavior near the $C_b$-$C_{dS}$ transition 
along a line in the $C_b$ phase at $k_0=0.75$ as in~\cite{lbrunning}:
while our estimation of the the critical parameter $\Delta_{crit}$ is compatible with 
the earlier result, we found that the critical index significantly differs
from the previous estimate.

We regard this last result as the most relevant of ours, as it points out that the use of the dual graph method might be questioned in some cases. 
Observing different critical indices in the approach to a phase transition line, indeed, implies that the length scales found by using the two methods cannot be \emph{simultaneously} compared with other length scales coming from different observables to look for a continuum limit.

While the FEM, extrapolated at high refinement levels, gives access to large scale properties of the full simplicial manifold, 
one big advantage of the dual graph approach is its relatively low computational cost. 
It might well be that in approaching a CDT continuum limit the two methods 
would show an agreement, or that their disagreement could be corrected by studying the scaling 
of some dimensionless parameters that may connect the length scales provided by the two methods. 
Nevertheless, we believe our warning should be taken into account by future studies, and the FEM should remain as a useful tool for CDT investigations.

The broadness of the Finite Element Method framework allows for extensions and variations
which we had no time and resources to consider in this work.
For example, the freedom in the choice of the refinement procedure can make 
for faster convergence, by using higher-order basis functions, 
or a mix of these with mesh refinements, or other types of FEM representations. 
Furthermore, in this work, we have only considered the application of FEM to the LB eigenproblem,
but the formalism is powerful enough to undertake more general tasks, 
like a properly defined representation of the introduction of new coupling terms
in the action, both gravitational ($f(R)$ extensions) and 
with matter and gauge fields 
(whose propagation properties would be unbiased by 
the geodesic overestimation of the dual graph representation), 
or the study of observables which have been limited in their definition by the attempts of 
embedding it in a dual graph structure, 
like the Wilson loop observable introduced in~\cite{cdt_Wilson_loop}. 
In general, having access to the true geodesics of the simplicial manifolds
(or at least to an arbitrarily good approximation of them), 
can also be useful, for example, to build and describe light cone observables, 
which may have, in principle, interesting phenomenological implications.

We plan to discuss the more challenging task of applying FEM techniques to 
the LB eigenproblem of full four-dimensional CDT configurations in a future work, 
as the refinement procedure requires more computational resources at higher dimensions.
Apart from the geodesic overestimation, for the analysis of four-dimensional CDT configuration 
there is also the problem that dual graph cannot faithfully represent the metrical properties of triangulations, but only their adjacency relations,
since the Wick rotated spacelike and timelike links have not, in general, the same length
(apart for $\Delta=0$), and we stress again that the Laplacian matrix of the dual graph has a clear relation with the LB operator only for equilateral triangulations.
Therefore, the FEM formalism (or any other formalism taking into consideration the anisotropy
in four-dimensional $4$-simplices) could be used instead, and this may have,
for example, important effects on the form of 
the dimensional reduction pattern~\cite{cdt_spectral_dim}.
We have also left out an analysis of the structure of FEM eigenvectors, 
which, together with the eigenvalues, contain complete information 
on the geometrical properties of the manifolds. 
A possible application of the eigenvectors is briefly discussed in the Appendix, 
where we define a Fourier transform of the local curvature observable, 
which can be useful, for example, to the construction of a smoothed curvature observable
(by truncating contributions from eigenvector associated to eigenvalues above a threshold).

\acknowledgments
 
We thank Massimo D'Elia for the useful comments and suggestions and Renate Loll and Jan Ambj{\o}rn for the stimulating discussions.
Numerical simulations have been performed on the MARCONI machine at CINECA,
based on the agreement between INFN and CINECA (under projects INF19\_npqcd and INF20\_npqcd),
while numerical analyses have been performed at the IT Center of the Pisa University.

\begin{appendices}

\appendix

\section{Technical details of the application of FEM to CDT simplicial manifolds}\label{sec:details}

In the following we explain in detail how we apply the FEM to our case of interest, the calculation of the LB eigenvalues on a simplicial manifold: in Section~\ref{subsec:fineigp} we show that each step of the method consists in solving the eigenproblem of a finite-dimensional, symmetric matrix; in Section~\ref{subsec:mat_elem} we explain how we calculate the matrix elements and solve this problem for each step; in Section~\ref{subsec:refinement} we outline the way we choose the sequence $\{\mathcal{V}_r\}_{r=0}^\infty$ of subspaces of $H^1$ we need for the application of the FEM, that in our case turns out to be a \textit{refinement procedure} for our simplicial manifold.

\subsection{LB eigenproblem in a FEM finite-dimensional subspace}\label{subsec:fineigp}

As stated in Section~\ref{sec:femintro}, in each step of the FEM the LB eigenproblem is reduced to the eigenproblem:
\begin{equation}\label{eq:weak-eigp-rep}
\int_\mathcal{M}\! d^dx \;\nabla \phi(\mathbf{x}) \nabla f(\mathbf{x}) = \lambda \int_\mathcal{M}\! d^dx \; \phi(\mathbf{x}) f(\mathbf{x}),
\end{equation}
in a generic finite-dimensional subspace $\mathcal{V}$. The method requires that, besides $f$, also the test functions are picked from this subspace: this means that the problem~\eqref{eq:weak-eigp-rep} reduces to a finite set of linear conditions.
Indeed, let $\{\phi_i\}_{i=1,\dots,N}$ be a basis of this subspace.
In this way, any function $f\in \mathcal{V}$ can be written as:
\begin{equation}
f(\mathbf{x})=\sum_{i=1}^{N} c_i \phi_i(\mathbf{x}).
\end{equation}
For each basis function $\phi_i$, equation~\eqref{eq:weak-eigp-rep} can be rewritten in the
form of a finite-dimensional generalized eigenvalue problem:
\begin{equation}\label{eq:fem-eigp}
L \vec{c} = \lambda M \vec{c}
\end{equation}
where we have introduced the two matrices $L$ and $M$ with matrix elements:
\begin{align}\label{eq:FEM_genL}
L_{i,j} &\equiv \bigintssss_{\mathcal{M}} \!d^d \mathbf{x}\,  \vec{\nabla}\phi_i(x) \cdot \vec{\nabla}\phi_j(x), \\[5pt]\label{eq:FEM_genM}
M_{i,j} &\equiv \bigintssss_{\mathcal{M}}\!d^d \mathbf{x}\, \phi_i(x) \phi_j(x).
\end{align}
Both matrices are symmetric, $M$ is positive-definite
($c^\top M c$ is nothing but the integral of the square of a function),
and $L$ is positive-semidefinite ($c^\top L c$ is the integral of the square gradient
of a function, then it can be zero if the constant function belongs to the subspace);
then this can be seen, after having inverted $M$,
simply as the eigenproblem of a positive-semidefinite matrix
in the following way: since $M$ is symmetric and positive,
it admits a (symmetric positive, then invertible) square root $M^{1/2}$
and Equation~\eqref{eq:fem-eigp} is equivalent to:
\begin{equation}
M^{-1/2}LM^{-1/2} (M^{1/2} c)= \lambda (M^{1/2} c),
\end{equation}
that is the eigenproblem of the symmetric non-negative matrix $M^{-1/2}LM^{-1/2}$
(non-negativity is obvious) with the eigenvectors simply read on a different
basis through the coordinate change $M^{1/2}$.
The eigenvectors $v=M^{1/2} c$ are orthonormal, so the vectors $c$ that solve
the problem~\eqref{eq:fem-eigp} are not orthonormal with respect to
the canonical scalar product but to that induced by $M$.

An interesting observation (see~\cite{fem_hughesbook,fem_strangbook}) is that the eigenvalues of
the finite-dimensional problem~\eqref{eq:fem-eigp} always overestimate
the exact LB eigenvalues $\lambda_n^{(exact)} \leq \lambda_n^{(FEM)}$,
so, we expect that the eigenvalues $\lambda_n^{(FEM,r)}$ obtained as solution to
the eigenproblems in a sequence of subspaces $\mathcal{V}_r\to H^1$, would converge
\textit{from above}, as indeed observed in the numerical results of Section~\ref{sec:examples}.

\subsection{Matrix elements and solution of the FEM eigenproblem}
\label{subsec:mat_elem}

In order to proceed, we need to choose our sequence of subspaces of $H^1$, calculate the respective matrix elements and solve the eigenproblem~\ref{eq:fem-eigp}.
As a starting point, we choose to restrict ourselves to the subspace $\mathcal{V}_0$ generated by
piecewise linear functions located at each of the vertices of the triangulation,
in such a way that every basis function $\phi_i$ has value $1$ on the vertex labeled $v_i$
and value $0$ on the border of the union of $d$-simplices to which the vertex $v_i$
belongs and outside this region\footnote{In algebraic topology,
the subcomplex made of the union of all the $d$-simplices containing
a given $k$-subsimplex $\sigma^{(k)}$ is called the \emph{closed star} of $\sigma^{(k)}$.
In our definition of basis functions the support of $\phi_i$ coincides with
the closed star of the vertex labeled as $v_i$.}.
The reason for this choice is that it implies that off-diagonal
elements $L_{ij}$ and $M_{ij}$, $i \neq j$, are non-zero if and only if
the vertices $i$ and $j$ are connected by a ($1$-D) link of the triangulation,
thus associating matrix elements and links;
as a result, the two matrices are sparse as in the case of the dual graph,
with the aforementioned benefits.\\
For the following steps, since we want to keep the matrices sparse,
we use subsequent enlargements of this subspace obtained by considering similarly
defined piecewise linear functions after having \textit{refined} the triangulation,
that is, having subdivided each simplex into smaller simplices
(notice that this procedure will produce some simplices not similar to any of the
starting ones, as argued below).
It is not hard to realize that this means that the size of
the matrices $L$ and $M$ grows exponentially,
thus raising what will turn out to be the main practical issue,
the long computational time needed to achieve convergence.

At this point, we need to calculate the general form of the matrix elements of $L$ and $M$
for these subspaces of functions.
We are required to perform integrals that extend on many simplices:
on every simplex to which the vertex belongs, for diagonal elements, and
on every simplex that shares the link $(i,j)$, for off-diagonal elements.
For generic simplices, the contributions to a given matrix element coming from
the integrals on each simplex are in general different;
for this reason, we find it convenient first to calculate the contributions of
the integrals on a single simplex (as functions of its geometrical characteristics)
to the matrix elements relative to each of its vertices and links,
and then sum up these contributions to build the matrix elements.

For future convenience, we will denote by $M_{ij}^{(\sigma)}$ and $L_{ij}^{(\sigma)}$ the respective
contributions to the $M$ and $L$ matrix elements integrated on a single $d$-simplex $\sigma$.
In order to actually compute these contributions,
we have to choose a chart for the simplex $\sigma$ such that we can represent the linear behavior
of the basis functions $\{ \phi_i \}$ in that chart;
For this purpose, we adopted an \emph{absolute barycentric coordinate system}
which we will now define.

Let us consider a $d$-simplex $\sigma$ where the vertices
are labeled by ${\{v_i\}}_{i=0,\dots,d}$.
We can always place the vertex $v_0$ to the origin $\vec{x}_0=\vec{0}$ of
an $\mathbb{R}^d$ chart and the other vertices to respective
cartesian coordinates ${\{\vec{x}_i\}}_{i=1,\dots,d}$,
such that $|l_{ij}|\equiv \norm{\vec{x}_i-\vec{x}_j}$ ($i\neq j$)
are the lengths of the links $l_{ij}$; these constraints uniquely define the coordinates up to
rotation and possibly permutation (if some links have the same length).
The absolute barycentric coordinates $\{\xi^i\}_{i=0,1,\dots,d}$ are subjected to the constraints
$\xi^i \geq 0$ and $\sum\limits_{i=0}^d \xi^i = 1$, so that
a generic point in the simplex cartesian chart
can be written as $\vec{p} = \sum_{i=1}^{d} \xi^i \vec{x}_i = A \vec{\xi}$,
where the matrix $A \equiv (x^\alpha_i)$ represents the linear map between
barycentric and cartesian charts
and $\vec{\xi}=(\xi_1,\dots,\xi_d)$\footnote{Notice that since $\vec{x}_0=\vec{0}$,
the vector $\vec{\xi}$ do not involve the barycentric coordinate $\xi_0$,
which is completely fixed by the constraint $\sum\limits_{i=0}^d \xi^i = 1$. Therefore,
$A$ is an invertible square matrix.}.\\
The reason why the barycentric chart is so useful is that the basis FEM functions $\phi_i$
with a linear bump on the $v_i$ vertex of the simplex $\sigma$
is simply $\phi_i(A\vec{\xi}) = \xi^i$,
where, again, $\xi_0=1-\sum_{1\leq i\leq d} \xi_i$ has to be considered
as a function of the independent variables $\xi_{1\leq i\leq d}$.
Therefore, the single simplex contributions to $L$ and $M$ matrix elements in
Equations~\eqref{eq:FEM_genL} and~\eqref{eq:FEM_genM} can be computed by
changing integration variables from cartesian to barycentric coordinates,
which maps the simplex $\sigma$ to the \emph{standard simplex}:
\mbox{$\sigma^\prime \equiv \{\vec{\xi}\in\mathbb{R}^{d+1} |
\sum\limits_{i=0}^{d}\xi^i=1, \xi^i \geq 0, i=0,\dots,d\}$}:

\begin{align}\label{eq:Mmat}
M_{ij}^{(\sigma)} = |A| \int_{\sigma^\prime} d^{d}\xi \;\; \xi_i \xi_j \;,
\end{align}
and
\begin{equation}\label{eq:Lmat}
\begin{gathered}
L_{ij}^{(\sigma)} = \int_{\sigma} d^{d}x \;\; \sum_\alpha \frac{\partial}{\partial x^\alpha}\phi_{i}(\vec{x}) \frac{\partial}{\partial x^\alpha}\phi_{j}(\vec{x}) = \\
= |A| \sum_{m,n=1}^{d} {[{(A^\top A)}^{-1}]}_{mn} \int_{\sigma^\prime} d^{d}\xi\;\; \frac{\partial}{\partial \xi^m}\xi^i \frac{\partial}{\partial \xi^n}\xi^j, \\
\end{gathered}
\end{equation}
where $|A|$ is the determinant of the linear application $A$ introduced above, and we used
the relation \mbox{$\frac{\partial}{\partial x^\alpha}
= \sum_m {(A^{-1})}_{\alpha}^{m} \frac{\partial}{\partial \xi^m}$}.

The integrals in the variables $\xi_1,\xi_2,\dots,\xi_d$ in
Equations~\eqref{eq:Mmat} and~\eqref{eq:Lmat} are completely independent on
the metric properties of the simplex $\sigma$,
while all the metric dependence is encoded in the $A$ matrix.
Notice also that $|A|$, the determinant of the linear map between the
standard simplex $\sigma^\prime$ and the original simplex $\sigma$,
is equal to their volume ratio $|A|=d!\; vol(\sigma)$.\\
Computing the straightforward integrals in
Equations~\eqref{eq:Mmat} and~\eqref{eq:Lmat} we obtain the following expressions for the
$M$ and $L$ matrix elements:
\begin{align}\label{eq:Mmat3}
M_{ij}^{(\sigma)} &= vol(\sigma) \frac{1+\delta_{i,j}}{(d+2)(d+1)}\; \forall i,j=0,\dots,d \;, \\
L_{00}^{(\sigma)} &= vol(\sigma) \sum_{m,n=1}^{d} {[{(A^\top A)}^{-1}]}_{mn} \; ,\\
L_{0i}^{(\sigma)} = L_{i0}^{(\sigma)} &= - vol(\sigma)\sum_{m=1}^{d} {[{(A^\top A)}^{-1}]}_{mi}
\;\forall i=1,\dots,d \;,\\
L_{ij}^{(\sigma)} &= vol(\sigma) {[{(A^\top A)}^{-1}]}_{ij} \;\forall i,j=1,\dots,d \;.
\end{align}

From its very definition, it is straightforward to show that
the matrix elements of $A^\top A$
are all the scalar products between the position vectors of the vertices
of $\sigma$ different from $v_0$:
$\vec{x}_i^\top \vec{x}_j = \xi_i^\top (A^\top A) \xi_j={(A^\top A)}_{ij}$; therefore,
by the cosine rule we obtain
\begin{equation}\label{eq:cosmatrix}
\begin{gathered}
{(A^\top A)}_{ij}=|l_{0i}| |l_{0j}| \cos(\beta_{ij})=\\
=\frac{1}{2}(|l_{0i}|^2+|l_{0j}|^2-|l_{ij}|^2) \; \forall i,j=1,\dots,d \;,
\end{gathered}
\end{equation}
where $\beta_{ij}$ is the angle (in the cartesian chart) between $\vec{x}_i$ and $\vec{x}_i$,
and $|l_{ij}|$ is the length of the link connecting the vertices of $\sigma$ labeled with $v_i$ and $v_j$.\\
Since $(A^\top A)$ is a positive-definite and Hermitian $d\times d$ matrix,
the fastest method to obtain the matrix elements of its inverse is to
first compute its \emph{Cholesky decomposition}~\cite{cholesky,numerical_recipes},
which returns the unique upper triangular matrix $A$ with positive diagonal elements,
and then invert $A$ by forward substitution.\\

Finally, the generalized eigenproblem~\eqref{eq:fem-eigp} can be numerically solved by means
of standard techniques on symmetric sparse matrices.
Due to its robustness and scalability, for our numerical results, we choose to employ the
Krylov-Shur algorithm~\cite{Krylov-Shur_algo} implemented
in the SLEPc library~\cite{slepc}.

As argued in Section~\ref{sec:femintro},
in order to obtain arbitrarily accurate estimates on the simplicial manifold $\mathcal{T}$,
we must build a sequence of approximating subspaces $\{\mathcal{V}_r(\mathcal{T})\}$
such that $\mathcal{V}_r(\mathcal{T})\subset\mathcal{V}_{r+1}(\mathcal{T})$
and $\mathcal{V}_r(\mathcal{T})\subset H^1(\mathcal{T})$ for all $r\geq 0$,
solve the eigenproblem on each subspace (up to a certain threshold $\bar{r}$),
and then extrapolate the results in the limit $r\to \infty$.

The procedure of building a subspace $\mathcal{V}_{r+1}(\mathcal{T})$ starting from
$\mathcal{V}_{r}(\mathcal{T})$ in a sequence with the properties stated above
is called \emph{refinement}, and $r$ is referred to as the \emph{refinement level}.
Our problem must be solved for such a number of refinement levels $\bar{r}$,
that the estimates of the eigenvalues reach convergence, which is signaled by a sufficiently
small relative variation between two subsequent estimates.
However, in general, whether the convergence has been reached or not may depend on
the order $n$ of the eigenvalue (or on the generic observable) into consideration.

\subsection{Refinement procedure}\label{subsec:refinement}

We argued that the initial subspace $\mathcal{V}_0(\mathcal{T})$ of piecewise-linear
functions on the original simplicial manifold $\mathcal{T}$
is just an approximation to $H^1(\mathcal{T})$.
There is a plethora of strategies which can be employed in order to obtain better approximations
of the full Hilbert space:
for instance, \emph{higher-order finite element methods} consist in using piecewise-polynomials
of maximal degree $r$ as approximating basis instead of the piecewise-linear ones;
this makes the approximating space $\mathcal{V}_r(\mathcal{T})\subset H^1(\mathcal{T})$
$r$ times bigger than its subset $\mathcal{V}_0(\mathcal{T})$,
and the results obtained more accurate (in particular, arbitrarily accurate for $r \to \infty$).
The very simple technique that we employ is called \emph{mesh refinement},
and consists in using again a basis of piecewise-linear bump functions,
but for a new triangulation $\mathcal{T}^{(r+1)}$
(i.e., $\mathcal{V}_r(\mathcal{T}) \equiv \mathcal{V}_0(\mathcal{T}^{(r)})$), where
the $d$-simplices of the triangulation in the previous
refinement level $\mathcal{T}^{(r)}$ have been partitioned into smaller $d$-simplices.

In general, the dimension of the approximating space grows as the number of new vertices;
however, this does not always imply convergence, i.e., not every sequence of approximating
spaces $\{\mathcal{V}_r\}$ with strictly increasing dimension is guaranteed to converge
to the infinite-dimensional Sobolev space $H^1(\mathcal{T})$.

\begin{figure}[h]
\centering
\begin{tikzpicture}[scale=0.5]
\tikzstyle{every node}=[draw,shape=circle,fill]
\foreach \shift in {0,5,10}
{\draw[xshift=\shift*1cm] (0,0) -- (4,0) -- (2,3.46) -- cycle;}
\newcommand*{\defcoords}{%
\coordinate (A1) at (2,3.46/3.0);
\coordinate (B1) at (2.0,0.0);
\coordinate (B2) at (1.0,3.46/2.0);
\coordinate (B3) at (3.0,3.46/2.0);
}

\begin{scope}
\defcoords
\node[scale=0.5] at (A1) {};
\draw (A1) -- (0,0);
\draw (A1) -- (4,0);
\draw (A1) -- (2,3.46);
\end{scope}

\begin{scope}[shift={(5cm,0)}]
\defcoords
\draw (B1) node[scale=0.5]{} -- (B2) node[scale=0.5]{} -- (B3) node[scale=0.5]{} -- cycle;
\end{scope}

\begin{scope}[shift={(10cm,0)}]
\defcoords
\node[scale=0.5] at (A1) {};
\node[scale=0.5] at (B1) {};
\node[scale=0.5] at (B2) {};
\node[scale=0.5] at (B3) {};
\draw (A1) -- (B1);
\draw (A1) -- (B2);
\draw (A1) -- (B3);
\draw (A1) -- (0,0);
\draw (A1) -- (4,0);
\draw (A1) -- (2,3.46);
\end{scope}

\node[draw=none,fill=none] at (2,-1) {A};
\node[draw=none,fill=none] at (7,-1) {B};
\node[draw=none,fill=none] at (12,-1) {C};

\end{tikzpicture}
\caption{Three possible types of refinements of a $2$-dimensional simplex.
The new vertices added are dotted.}
\label{fig:ref2d_types}
\end{figure}

For example, three possible partitions of a triangle in a two-dimensional triangulation are shown
in Figure~\ref{fig:ref2d_types}: the A type of refinement, even if iterated an infinite number of times
on its subsimplices, cannot represent functions with generically varying value on
the original links, while this is possible for both the B and C types of
refinement\footnote{It is even possible to mix different refinement strategies: for example,
alternating type A, B and C refinements from Figure~\ref{fig:ref2d_types}
could be useful.}.
In general, the convergence of a sequence of refinements is
guaranteed whenever the maximum of the diameters of the elements vanishes at infinite refinement
level~\cite{fem_hughesbook,fem_strangbook} (this does not hold for the type A refinement).
There are many ways to refine a triangulation,
because there is much freedom in the choice of positions for the new vertices and shapes for the
subsimplices\footnote{One could also allow for non-simplicial elements like bounded
convex polytopes, but the expressions for the $L$ and $M$ matrix elements would be overly
complicated without particular advantages, so we will always consider
simplicial elements in our discussions}.

In order to reach a better convergence rate,
minimizing the maximum link lengths (and therefore the simplex diameters) at each step,
it is customary to refine by adding new vertices on the center of the links of
the previous iteration, connecting all of them with new links,
and then filling up the remaining space; in two dimensions,
this procedure corresponds to the refinement type shown as B in Figure~\ref{fig:ref2d_types},
where the remaining space is another (upside down) triangle
and therefore the original triangle is partitioned into $4$ triangles.
Starting from a two-dimensional triangulation made of equilateral triangles,
this type of refinement produces another triangulation with $4$ times the number of triangles,
and that these ones are still equilateral and all with the same sides
(halved with respect to the previous ones).
This observation will be useful in Section~\ref{sec:examples},
since it makes it possible to build a refinement procedure also for dual graphs of
two-dimensional triangulations, and which allows us to compare both FEM and dual graph methods
in a regime where both would converge to the exact LB spectrum.
As we will discuss in a moment, preserving the regularity of the subsimplices in the partitions
is actually impossible in dimensions higher than two.
Another interesting fact that we would like to point out about this type of refinement is that,
adding a vertex in the middle of a preexisting link (as for the types of refinement
B and C in Figure~\ref{fig:ref2d_types}),
forces a partitioning also on neighboring simplices, so that is not possible, for example,
to refine only a certain region of the triangulation, but this process has to occur
globally\footnote{Except for refinements as the type A in Figure~\ref{fig:ref2d_types},
which alone, however, do not guarantee any convergence.}.

In dimensions higher than two, this procedure becomes more complicated, as we will now argue.
Let us consider a single vertex $v$ of a $d$-simplex $\sigma$,
which is connected to $d$ links of $\sigma$.
When we put new vertices $v_1^\prime,v_2^\prime,\dots,v_d^\prime$ in the middle of
these links (and connect them with new links),
we automatically obtain a subsimplex with $v$ and the $v_{1\leq j \leq d}^\prime$ as vertices.
In this way, the original simplex $\sigma$ is partitioned
into $d+1$ subsimplices (one for each vertex),
\textit{plus} the remaining region of space left inside $\sigma$;
the shape of this region is a polytope
called the \emph{rectification} of the $d$-simplex $\sigma$,
or also \emph{critical truncation} of $\sigma$ (see~\cite{coxeter2012regular}
for more details), and it corresponds to a genuine $d$-simplex only in two dimensions
(i.e., the inner triangle between new vertices in type B of Figure~\ref{fig:ref2d_types}).

\pgfmathsetmacro{\a}{0.85}
\pgfmathsetmacro{\m}{\a*sqrt(3)/2}
\pgfmathsetmacro{\H}{sqrt(1-\m^2/4)}
\pgfmathsetmacro{\h}{sqrt(\H^2-\m^2/4)}
\begin{figure}[h]
\centering
\begin{tikzpicture}[scale=1.4,
font=\footnotesize,
x={({cos(33)*2cm}, {-sin(33)*2cm})},
y={({cos(25)*2cm}, {sin(25)*2cm})},
z={(0cm,1.7cm)},
]

\begin{scope}
\coordinate (T0) at (0,0,0);
\coordinate (T1) at (\a,0,0);
\coordinate (T2) at (0,\a,0);

\draw[] (T0) -- (T1) -- (T2) --cycle;

\path[] (T0) -- ($(T1)!0.5!(T2)$) coordinate(Ha);
\path[] (T1) -- ($(T0)!0.5!(T2)$) coordinate(Hb);
\coordinate[] (M) at (intersection of T0--Ha and T1--Hb);

\path[] (M) --+ (0,0,\H) coordinate (T3);
\draw[] (T0) -- (T3) ;
\draw[] (T1) -- (T3) ;
\draw[] (T2) -- (T3) ;

\draw[] ($(T0)!0.5!(T1)$) node[shape=circle,fill,scale=0.5]{} coordinate (T01);
\draw[] ($(T0)!0.5!(T2)$) node[shape=circle,fill,scale=0.5]{} coordinate (T02);
\draw[] ($(T1)!0.5!(T2)$) node[shape=circle,fill,scale=0.5]{} coordinate (T12);
\draw[] ($(T0)!0.5!(T3)$) node[shape=circle,fill,scale=0.5]{} coordinate (T03);
\draw[] ($(T1)!0.5!(T3)$) node[shape=circle,fill,scale=0.5]{} coordinate (T13);
\draw[] ($(T2)!0.5!(T3)$) node[shape=circle,fill,scale=0.5]{} coordinate (T23);

\draw (T01) -- (T02);
\draw (T01) -- (T03);
\draw (T01) -- (T12);
\draw (T01) -- (T13);
\draw (T02) -- (T03);
\draw (T02) -- (T12);
\draw (T02) -- (T23);
\draw (T03) -- (T13);
\draw (T03) -- (T23);
\draw (T12) -- (T13);
\draw (T12) -- (T23);
\draw (T13) -- (T23);
\draw (T13) -- (T23);

\draw[] ($(T01)!0.5!(T23)$) node[shape=circle,fill,scale=0.5]{} coordinate (Tc);

\draw[] (T01) -- (Tc);
\draw[] (T02) -- (Tc);
\draw[] (T03) -- (Tc);
\draw[] (T12) -- (Tc);
\draw[] (T13) -- (Tc);
\draw[] (T23) -- (Tc);

\node[draw=none,fill=none,scale=1.1] at (1cm,-1.1cm) {with central vertex};
\end{scope}

\begin{scope}[shift={(3cm,0)}]
\coordinate (T0) at (0,0,0);
\coordinate (T1) at (\a,0,0);
\coordinate (T2) at (0,\a,0);

\draw[] (T0) -- (T1) -- (T2) --cycle;

\path[] (T0) -- ($(T1)!0.5!(T2)$) coordinate(Ha);
\path[] (T1) -- ($(T0)!0.5!(T2)$) coordinate(Hb);
\coordinate[] (M) at (intersection of T0--Ha and T1--Hb);

\path[] (M) --+ (0,0,\H) coordinate (T3);
\draw[] (T0) -- (T3) ;
\draw[] (T1) -- (T3) ;
\draw[] (T2) -- (T3) ;

\draw[] ($(T0)!0.5!(T1)$) node[shape=circle,fill,scale=0.5]{} coordinate (T01);
\draw[] ($(T0)!0.5!(T2)$) node[shape=circle,fill,scale=0.5]{} coordinate (T02);
\draw[] ($(T1)!0.5!(T2)$) node[shape=circle,fill,scale=0.5]{} coordinate (T12);
\draw[] ($(T0)!0.5!(T3)$) node[shape=circle,fill,scale=0.5]{} coordinate (T03);
\draw[] ($(T1)!0.5!(T3)$) node[shape=circle,fill,scale=0.5]{} coordinate (T13);
\draw[] ($(T2)!0.5!(T3)$) node[shape=circle,fill,scale=0.5]{} coordinate (T23);

\draw[] (T01) -- (T02);
\draw[] (T01) -- (T03);
\draw[] (T01) -- (T12);
\draw[] (T01) -- (T13);
\draw[] (T02) -- (T03);
\draw[] (T02) -- (T12);
\draw[] (T02) -- (T23);
\draw[] (T03) -- (T13);
\draw[] (T03) -- (T23);
\draw[] (T12) -- (T13);
\draw[] (T12) -- (T23);
\draw[] (T13) -- (T23);
\draw[] (T13) -- (T23);

\draw[] (T01) -- (T23);

\node[draw=none,fill=none,scale=1.1] at (1cm,-1.1cm) {without central vertex};
\end{scope}

\end{tikzpicture}
\caption{Rectification of the tetrahedron (inner octahedron)
and two types of octahedral partitions:
with central vertex (left) and without central vertex (right).
The new vertices added are dotted.}
\label{fig:ref3d}
\end{figure}
In three dimensions, the rectification of a $3$-simplex produces an octahedron,
which has to be partitioned into tetrahedra.
The simplest, most symmetric way to partition this octahedron into $3$-simplices is to create
a new vertex in its center, and connect it to the vertices of its faces in order to
form $8$ new tetrahedra,
as shown in the left of Figure~\ref{fig:ref3d}.
As another possible refinement procedure,
we can also choose not to add the central vertex on the octahedron
in the center of each simplex, but,
instead, to add only a diagonal link between two of its antipodal
vertices\footnote{Chosing the largest diagonal as new link turns out to be
the optimal choice, since it minimize that maximum element diameter,
and ensure a better convergence.},
as shown in the right of Figure~\ref{fig:ref3d},
so that the number new tetrahedra into which it can be partitioned become $4$
(the ones around the selected diagonal).
This choice makes the subspace dimension grow at increasing refinement level with a slightly
slower rate than with the addition of a central point,
and without substantial loss in accuracy.

The refinement procedure of four-dimensional triangulations will not be used in this work,
since its aims are to present the method and show a comparison with
previous data obtained with spectral analysis of
dual graphs of spatial slices~\cite{lbstruct,lbrunning},
which requires only three-dimensional refinements.
Moreover, spectral analysis of four-dimensional triangulations becomes
computationally demanding at higher refinement levels,
due to its large rate of growth of
the subspace dimension\footnote{It may be possible to mitigate the computational efforts
by using a multiscale technique,
where the eigenspaces found for a triangulation at a certain refinement level are used
as ansatz for the next refinement level.}.
For these reasons, we will investigate
full four-dimensional triangulations with due care in a future work.
Nevertheless, for completeness purposes,
in the rest of this section we will briefly discuss which refinement procedures
are possible for four-dimensional triangulations.

The rectification of a $4$-simplex is more complicated than for lower dimensions,
since it does not produce a regular polytope, but what is usually called \emph{rectified 5-cell},
whose faces are $5$ regular tetrahedra and $5$ octahedra.
Again, adding a new vertex at the center of the inner rectified $5$-cell, and one new vertex
at the center of each octahedral face
(as in the first of the refinement strategies discussed above in three dimensions),
it is possible to symmetrically partition the original $4$-simplex into $50$ new $4$-simplices.
As in the three-dimensional case, it is also possible to not add a central vertex to the $5$ inner
octahedral faces, by splitting anisotropically them into $4$ new $3$-simplices each,
making the total counting of $4$-simplices of the partition of the original one add up to $30$.
This still amounts to a fast rate of growth of the Hilbert space dimension for increasing
refinement level, but this rate is definitely slower
(therefore better for the computational cost) than the one obtained with the addition
of central vertices on the octahedral faces of rectified $5$-cells,
and comes with no substantial loss in accuracy.
It is also possible to anisotropically partition the rectified $5$-cell without the addition of
an inner central point, but this becomes overly complex and not particularly helpful.

\section{Curvature observables in FEM}\label{sec:curvobs}

In the FEM framework, it is not hard to introduce a very useful new tool to study
the curvature of CDT simplicial manifolds, that may hopefully help to identify
unknown properties of the various phases of CDT: 
the ``Fourier transform'' of the scalar curvature $R$.\\
The eigenvectors of the LB operator are a complete basis, 
then any function can be decomposed in a superposition of them, 
with Fourier coefficients given by the scalar products between the function 
and the LB eigenvectors. 
The Fourier coefficients of $R$ contain information on its overall distribution, 
most of all on its typical scales.\\
As the approximate eigenvectors we find at each refinement step converge to the real ones, 
to find the Fourier coefficients it is enough to know how to calculate the scalar products 
between $R$ and the approximate eigenvectors.\\
Since the curvature has support on $(d-2)$-simplices and the discretized version of 
the integral of $R$ over the whole manifold is 
$\sum_{\sigma^{d-2}} 2\varepsilon_{\sigma^{d-2}} V_{\sigma^{d-2}}$, 
the most natural way to write the curvature relative to each $(d-2)$-simplex 
in a functional form is:
\begin{equation}
R(\sigma^{d-2})(\mathbf{x}) = 2\varepsilon_{\sigma^{d-2}} \delta(x_1-x_{1,\sigma^{d-2}})\delta(x_2-x_{2,\sigma^{d-2}}),
\end{equation}
where the variables in the Dirac deltas are to be identified, 
for each of the $d$-simplices that share the $(d-2)$-simplex, 
with two coordinates that run orthogonally to the $(d-2)$-simplex in local coordinates 
relative to a $d$-simplex. 
In simpler words, the integration of the $L^2$-scalar product, 
when the Fourier transform is computed, 
has to be performed only on the restricted domain given by the union of $(d-2)$-simplices.
The whole function $R(\mathbf{x})$ is then nothing but the sum of these contributions over 
the $(d-2)$-simplices.\\
Given that the approximate LB-eigenvectors found with the FEM are of the form:
\begin{equation}
u_n(\mathbf{x}) = \sum_{i=1}^{N_0} u_{n,i} \phi_{i}(\mathbf{x}),
\end{equation}
where the $\phi$s are the functions of that step's basis, 
it is not hard to perform the scalar products, 
obtaining the simple expressions we show in the following.

\subsection{Case d=2} 
In two dimensions the scalar curvature lives on 
the vertices of the triangulation, so the part left after integrating out 
the Dirac deltas is particularly simple to evaluate 
(remembering that $\phi_{i}$ has value $1$ on the vertex $i$ and $0$ on the others):
{\small
\begin{equation}
\hat{R}(n)=\int_{\mathcal{M}} \sum_{i=1}^{N_0} u_{n,i} \phi_{i}(\mathbf{x}) \sum_{j=1}^{N_0} R_j(\mathbf{x}) d^2x=\sum_{i=1}^{N_0} 2 \varepsilon_i u_{n,i}.
\end{equation}}
	
\subsection{Case d=3} 
In three dimensions, $R$ is associated with 1-D links; 
so, after integrating out the Dirac deltas, 
we are left with integrals in one dimension on the links:
\begin{equation}
\begin{gathered}
\hat{R}(n)=\int_{\mathcal{M}} \sum_{i=1}^{N_0} u_{n,i} \phi_{i}(\mathbf{x}) \sum_{j=1}^{N_1} R_j(\mathbf{x}) d^3x=\\
= \sum_{i=1}^{N_0} u_{n,i} \sum_{j| i\in l_j} \int_{l_j} 2 \varepsilon_j \phi_i(\mathbf{x}) dx,
\end{gathered}
\end{equation}
where the summations have been simplified thanks to the property of each $\phi_i$ 
of being $0$ outside the simplices sharing the vertex $i$.
Considering also their piecewise linearity, 
the integrals left are easily evaluated to $\frac{1}{2}meas(l_j)$ 
(it is simply the area of a triangle), yielding for the total Fourier components:
\begin{equation}
\hat{R}(n)=\sum_{i=1}^{N_0} u_{n,i} \sum_{j| i\in l_j} \varepsilon_j meas(l_j).
\end{equation}
	
\subsection{Case d=4} 
In four dimensions the scalar curvature has support on the 
2-simplices (triangles); so, after integrating out the Dirac deltas, 
there remain two-dimensional integrals on the triangles:
\begin{equation}
\begin{gathered}
\hat{R}(n)=\int_{\mathcal{M}} \sum_{i=1}^{N_0} u_{n,i} \phi_{i}(\mathbf{x}) \sum_{j=1}^{N_2} R_j(\mathbf{x}) d^4x=\\
= \sum_{i=1}^{N_0} u_{n,i} \sum_{j| i\in T_j} \int_{T_j} 2 \varepsilon_j \phi_i(\mathbf{x}) d^2x,
\end{gathered}
\end{equation}
where the summations have been reduced in a similar way to the previous case. 
Again, the piecewise linearity of the $\phi$s allows us to straightforwardly evaluate 
the integrals to $\frac{1}{3}meas(T_j)$ (this time it is the volume of a pyramid), 
thus obtaining for the Fourier components:
\begin{equation}
\hat{R}(n)=\sum_{i=1}^{N_0} u_{n,i} \sum_{j| i\in T_j} \frac{2}{3}\varepsilon_j meas(T_j).
\end{equation}

Notice that the generalization is straightforward: in arbitrary dimension 
for each vertex $i$ one is left with a summation of integrals on the $(d-2)$-simplices 
sharing $i$, that every time evaluate to $\frac{1}{d-1}meas(\sigma^{d-2}_j)$. 
This corresponds to the (a priori) naive idea of considering each $\phi_i$ having ``support'' 
only on vertex $i$ and redistribute the (integrated) curvature associated 
to each $(d-2)$-simplex in equal parts between its $(d-1)$ vertices.
Linearity is what ensures this works.\\
The expressions we have found for the Fourier coefficients do not represent 
a great further computational effort once the problem 
in Equation~\eqref{eq:fem-eigp} has been solved, 
then from this point of view it would be no problem to include them into 
the analysis tools we use in CDT, once it will have become clear how to build 
truly physically meaningful curvature observables 
out of them. 

\end{appendices}

\end{document}